\documentclass[aps,prl,reprint,superscriptaddress,nofootinbib,longbibliography]{revtex4-1}

\usepackage{amsmath,amssymb,bm}
\usepackage{graphicx}
\usepackage{booktabs}
\usepackage{array}
\usepackage{xcolor}
\usepackage[colorlinks=true,linkcolor=blue,citecolor=blue,urlcolor=blue,bookmarks=false]{hyperref}

\graphicspath{{figures/}}

\newcommand{\tp}{t'}

\newcommand{\epsk}{\varepsilon_{\mathrm{NN}}(\mathbf{k})}
\newcommand{\wzero}{\omega_{\mathrm{GZ}}}
\newcommand{\ReG}{\mathrm{Re}\,G}
\newcommand{\ImG}{\mathrm{Im}\,G}
\newcommand{\rhoG}{\rho_G}
\newcommand{\nk}{n(\mathbf{k})}
\newcommand{\nmix}{n_{\mathrm{mix}}}

\newcommand{\beq}{\begin{eqnarray}}
\newcommand{\eeq}{\end{eqnarray}}

\newcommand{\disp}[1]{Eq.~(\ref{#1})}
\newcommand{\Disp}[1]{Equation~(\ref{#1})}

\newcommand{\refdisp}[1]{Ref. [\onlinecite{#1}]}
\newcommand{\figdisp}[1]{Fig. \ref{#1}}
\newcommand{\Figdisp}[1]{Figure \ref{#1}}

\begin{document}

\title{Green-function Zeros Encode Competing Mott and Charge-ordering Scales}

\author{Peizhi Mai}
\email{peizhimai@gmail.com}
\affiliation{Department of Physics and Astronomy, The University of Tennessee, Knoxville, Tennessee 37996, USA}
\affiliation{Department of Physics and Institute of Condensed Matter Theory, University of Illinois at Urbana--Champaign, Urbana, Illinois 61801, USA}
\author{Philip W. Phillips}
\email{dimer@illinois.edu}
\affiliation{Department of Physics and Institute of Condensed Matter Theory, University of Illinois at Urbana--Champaign, Urbana, Illinois 61801, USA}
\date{\today}

\begin{abstract}
While correlated insulators are devoid of low-energy quasiparticle poles,  their Green functions retain clean momentum structure through zeros.  However, precisely what zeros imply is not clear.  By studying the extended Hubbard model both analytically and numerically, we establish a new paradigm for strongly correlated matter:  the dispersion of Green function zeros is determined by both microscopic spin-spin correlations and defect kinematics.  In fact, we find that the dispersion changes discontinuously across the transition between the  Mott and the checkerboard charge-density wave phases as is expected for a first-order transition. 
This physics is robust to the inclusion of further neighbor hopping which simply fine tunes the spin-correlation or charge-defect kinematics.  We conclude that it is the {\it dispersion} of the Green-function zeros that encode the physics of ordering resultant from the strong correlations. 
\end{abstract}

\maketitle

\section{Introduction}

Since their inception, Mott\cite{mott} insulators have always been associated with ordering. This is natural because a half-filled band with spin-1/2 electrons in the square lattice will order antiferromagnetically at sufficiently low temperature.  However, as Anderson\cite{anderson} has emphasized, the ordering that ensues is a consequence of the strong correlations not the root cause of the insulating state.  Precisely how to disentangle ordering from the insulating state has been the persistent question of Mott insulation.   In 2003,  Dzyaloshinskii\cite{dzyaloshinskii2003zeros} pointed out that zeros,  of the single-particle Green function can obtain in a Mott insulator.
In essence, zeros arise from a balancing act between spectral weight above and below the chemical potential.  While poles count quasiparticles, it is unclear what zeros entail.  Nonetheless, because cuprate superconductors start off as Mott insulators, the precise meaning of zeros is a hotly debated topic\cite{rosch,stanescu2007luttinger,sakai2009polezero,dave2013absence,gurarie2011greens,seki2017topological,fabrizio2022luttinger,skolimowski2022luttinger,blason2023unified,stefan,setty2024electronic,milliszeros}.  A point of concurrence among the different approaches is that zeros emerge from a divergence of the self energy and hence lie outside perturbation theory. Such breakdowns can lead to non-analyticities in the Luttinger-Ward functional which can generate a triangle-diagram anomaly\cite{chubukov,lanave} in the Luttinger count.  

In general, the zero-crossings will depend on momentum and frequency. Hence, what is needed to delineate the full zeros spectrum (not just the zero-frequency crossing which defines the Luttinger-Ward\cite{lw} surface) is a sufficiently non-perturbative method that has momentum and real-frequency resolution.  Consequently, methods that require analytic continuation and are susceptible to the sign problem are difficult to resolve zero crossings.  A few of the methods employed include numerics\cite{senechal,stanescu2006hiddenzeros} and diagrammatic expansions\cite{stepanov2024interconnected} which yield a dispersion of the form $\epsilon_{k+Q}/U^2$ where $\epsilon_{k+Q}$ is the bare band energy shifted by the M-point momentum and $U$ is the on-site repulsion.  A distinct feature of this dispersion is that it vanishes in the $U=\infty$ limit.  That is, the zeros do not disperse when the correlations are infinitely strong, reminiscent of the dynamical mean-field theory (DMFT)\cite{georges1996dmft} result of dispersionless zeros at $\omega=0$.   A study of the Su-Schrieffer-Heeger-Hubbard\cite{lehman} model finds spin-spin correlations play a role in determining the magnitude of the zero dispersion, whereas the band Hatsugai-Kohmoto\cite{hatsugai1992exactly,sc} (HK) model yields a dispersing band of zeros determined solely by the non-ineracting $\varepsilon(\mathbf{k})$.  

\begin{figure}[!t]
\centering
\includegraphics[width=\columnwidth]{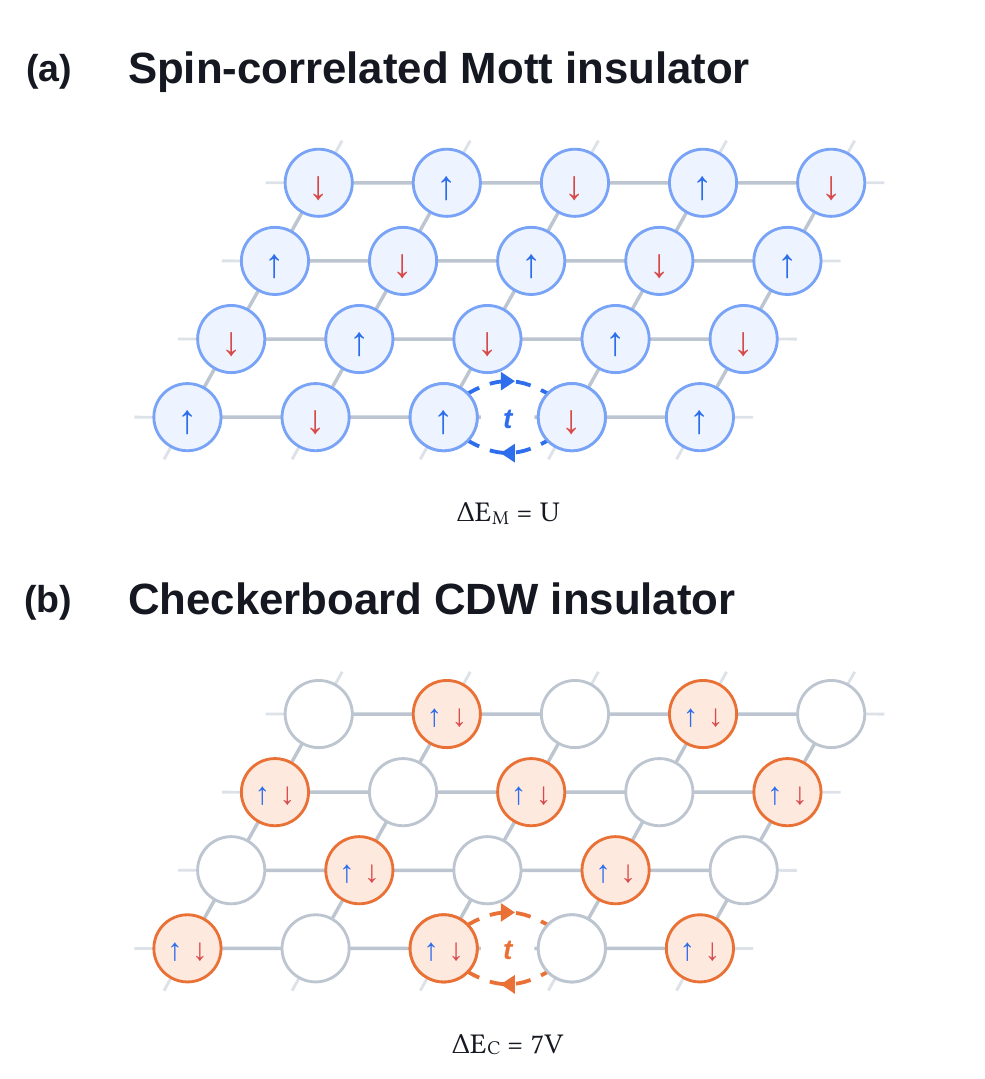}
\caption{\textbf{Distinct correlated insulators.} (a) In the spin-correlated Mott insulator, sites are predominantly singly occupied, and NN hopping $t$ virtually accesses a doublon--holon configuration at energy cost $\Delta E_{\mathrm M}=U$ with $V=0$. (b) In the checkerboard CDW insulator, doubly occupied and empty sites alternate, and the corresponding virtual hop accesses two singly occupied defects at cost $\Delta E_{\mathrm C}=7V$  with $U=0$. Dashed counterpropagating arrows identify the virtual process on a representative bond; the finite patches illustrate the local backgrounds.}
\label{fig:real-space-backgrounds}
\end{figure}
To arrive at a general conclusion regarding zeros, we focus on a model that has two competing ordering tendencies,  Mott antiferromagnetism and charge-density-wave (CDW) ordering shown in Fig. (\ref{fig:real-space-backgrounds}).  Hence our conclusions apply beyond Mott physics.  Our tool is a generalization of the  recently developed\cite{mai2026twisting} momentum-mixing Hatsugai-Kohmoto (MMHK) framework to include nearest-neighbor (NN) repulsion $V$, enabling a dynamical study of the competition between spin-correlated Mott and checkerboard CDW states.   As demonstrated\cite{mai2026twisting} previously, this framework interpolates systematically between the exactly solvable HK limit \cite{hatsugai1992exactly} and the Hubbard model while preserving the HK fixed-point structure \cite{fixedpoint}. What we explore here is the here-to-fore unearthed real-frequency spectral evolution. 

The key result we derive is a dispersion (not just the zero-frequency crossing discussed by Dzyaloshinskii\cite{dzyaloshinskii2003zeros}) relationship for the gap-interior zeros across the Mott--CDW transition.  Quite generally, we show that the simple relationship $\wzero(\mathbf{k})\simeq s_{\mathrm{NN}}\epsk$ captures the dispersion of the gap-interior zeros.  Here $s_{\rm NN}$ is a numerical factor determined by the strong-correlation spin or charge ordering physics and $\epsk$ is the non-interacting band dispersion.  Because on the Mott side, $s_{\rm NN}\approx-1.4$, whereas on the CDW side, $s_{\rm NN}=-8/7$, we find that the bandwidth of the zeros remains finite  in the infinite-interaction limit in contrast to the diagrammatic extension of DMFT\cite{stepanov2024interconnected} results discussed earlier.  Local spin correlations are absent from band HK, resulting in $s_{\rm NN}=1$.  The fact that in both MMHK and HK the width of the zero dispersion remains finite points to the underlying $Z_2$ universality\cite{z2}. The value of $s_{\mathrm{NN}}$ on both phases is verified numerically and analytically. The discontinuous jump in $s_{\mathrm{NN}}$ and the independent discontinuities in the charge and spin correlation establish the first-order nature of the Mott--CDW transition. Hence, we arrive at the surprising result that the GZ dispersion is determined by the underlying order that accompanies Mott--CDW physics.  Consequently, it is dispersive zeros that complete the Mott story and distinguish it from perturbative Slater physics.

\section{$V$-extended MMHK model and GZ dispersion}

We first define the $V$-extended MMHK construction at a general momentum-mixing resolution. We define $\mathcal C_{\nmix}=\{\mathbf r_\alpha\}_{\alpha=1}^{\nmix}$ as an $\nmix$-orbital mixing cell. In reciprocal space, translations of the reduced Brillouin zone (${\rm rBZ}_{\nmix}$) by the $\nmix$ discrete mixing momenta $\mathbf K$ tile the full Brillouin zone. The continuous vector $\mathbf f\in{\rm rBZ}_{\nmix}$ selects the same relative point within every tile, so a fixed-$\mathbf f$ sector contains the $\nmix$ physical momenta $\mathbf k=\mathbf K+\mathbf f$.  The operator $c_{\mathbf f\alpha\sigma}$ annihilates a spin-$\sigma$ fermion in the orbital $\alpha$ with momentum $\bf f$. The $V$-extended $\nmix$-MMHK Hamiltonian is
\begin{align}
H_{\mathrm{MMHK}}^{(\nmix)}
={}&\sum_{\mathbf f\in{\rm rBZ}_{\nmix}}
\Bigg[\sum_{\alpha,\beta,\sigma}
g_{\alpha\beta}^{t,\tp}(\mathbf f)\,
c^\dagger_{\mathbf f\alpha\sigma}c_{\mathbf f\beta\sigma}-\mu\sum_{\alpha\sigma }n_{\mathbf f\alpha\sigma} \nonumber\\[-0.15em]
&+U\sum_\alpha n_{\mathbf f\alpha\uparrow}n_{\mathbf f\alpha\downarrow}
+V\sum_{\langle\alpha\beta\rangle}n_{\mathbf f\alpha}n_{\mathbf f\beta}
\Bigg] .
\label{eq:model}
\end{align}
Here $g_{\alpha\beta}^{t,\tp}(\mathbf f)
=\sum_{\Delta\mathbf R}
t_{\alpha\beta}(\Delta\mathbf R)\,
e^{i\mathbf f\cdot(\mathbf r_\alpha-\mathbf r_\beta+\Delta\mathbf R)}$ and $t_{\alpha\beta}(\Delta\mathbf R)$ contains the NN and next-nearest-neighbor (NNN) hopping. The bond sum $\langle\alpha\beta\rangle$ contains each NN bond of the periodic internal mixing cell once. Crucially, the interaction operators in \disp{eq:model} carry the same sector momentum $\mathbf f$ as the kinetic operators; suppressing that label would obscure the defining MMHK decomposition. For $V=0$ and $\nmix=1$, \disp{eq:model} reduces to the solvable HK limit. Increasing $\nmix$ restores progressively more interaction-induced momentum-transfer processes. At fixed lattice size $N$, $\nmix=N$ makes the reduced zone a single point and exactly recovers the Hubbard interaction; the same completeness argument applied to the NN form factor $\mathcal V(\mathbf q)=2V(\cos q_x+\cos q_y)$ recovers $V\sum_{\langle ij\rangle}n_i n_j$. The full $V$-extended construction and its extended-Hubbard limit are derived in Supplemental Material, Sec.~1 \cite{mai2026twisting}.

This manuscript uses $\nmix=16$ with a $4\times4$ mixing cell, $\alpha\equiv\mathbf r=(x,y)$ and $x,y=0,1,2,3$. The corresponding scattering momenta are
$\mathcal K_{16}=\left\{(\pi/2)(m,n):m,n=0,1,2,3\right\}$. For a continuous sector offset $\mathbf f$, the physical momentum is $\mathbf k=\mathbf K+\mathbf f$ ($\mathbf K\in\mathcal K_{16}$ and $\mathbf f\in{\rm rBZ}_{16}$). 
After Fourier transforming the internal coordinate, we see that the hopping term in \disp{eq:model} yields a bare dispersion $\varepsilon(\mathbf k)=\varepsilon_{\mathrm{NN}}(\mathbf k)+\varepsilon_{\mathrm{NNN}}(\mathbf k)$, where $\varepsilon_{\mathrm{NN}}(\mathbf k)=-2t(\cos k_x+\cos k_y)$ and $\varepsilon_{\mathrm{NNN}}(\mathbf k)=-4\tp\cos k_x\cos k_y$ are the NN and NNN hopping contributions, respectively. Continuous $\mathbf f$ supplies thermodynamic momentum resolution, whereas $4\times4$ specifies the interaction-mixing resolution. We work at half filling with $\mu=U/2+4V$. We first set $\tp=0$ to establish the simple GZ-dispersion relation under strong correlations in the particle--hole-symmetric case. We then turn on $\tp$ to generalize it to a generic band without particle--hole symmetry, for which the GZ dispersion separates into distinct hopping contributions.

The central observable is the continuous low-spectral-weight solution to
\begin{equation}
  \ReG\!\left[\mathbf{k},\wzero(\mathbf{k})\right]=0,
  \; \wzero(\mathbf{k})\ \text{inside the central gap}.
  \label{eq:zero-definition}
\end{equation}
$\wzero(\mathbf{k})$ denotes the {\bf GZ dispersion}. Finite numerical broadening leaves a small residual $\ImG$ inside the gap that vanishes as the broadening is removed. In the zero-broadening limit, the $\omega=0$ intersection of this dispersion---the set of momenta satisfying $\wzero(\mathbf{k})=0$---is the conventional Luttinger surface $G(\mathbf{k},0)=0$.  A previous study\cite{stanescu2007luttinger} demonstrated that the Luttinger surface is pinned at the Fermi surface of the non-interacting system in the presence of particle-hole symmetry. It is the general behaviour of $\omega_{\rm GZ}$ which is not necessarily pinned to the Luttinger surface that we calculate here.

\section{Results}

\subsection{Benchmark validation}

\begin{figure}[!t]
\centering
\includegraphics[width=\columnwidth]{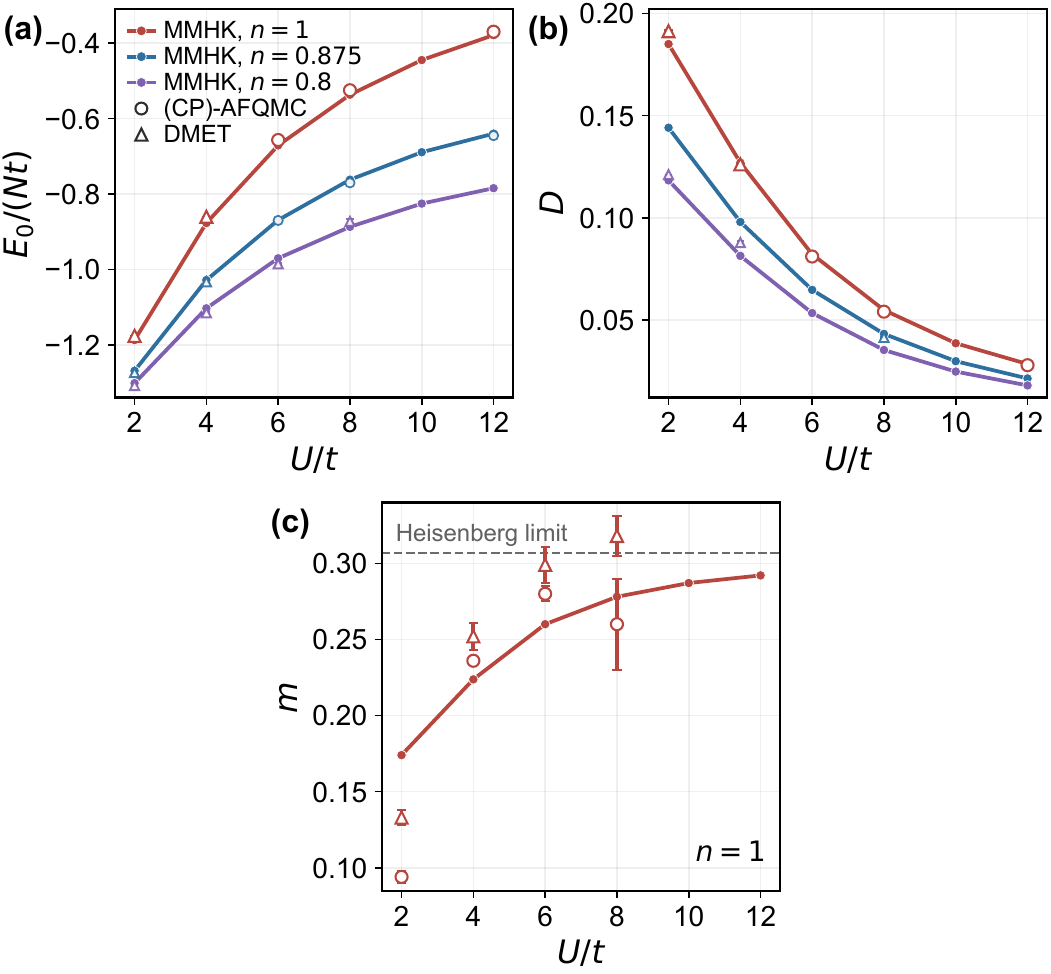}
\caption{\textbf{Benchmark validation of $16$-MMHK.} (a, b) Ground-state energy and double occupancy for $n=1$, $0.875$, and $0.8$. Color identifies density; filled circles connected by lines are MMHK, open circles are AFQMC or constrained-path AFQMC, and open triangles are DMET. (c) Half-filled staggered magnetization; the Heisenberg limit is shown. Benchmark-source selection, pointwise values, deviations, and source identities are reported in the Supplemental Material, Sec.~2 and Table~S1.}
\label{fig:benchmark-density}
\end{figure}

We first benchmark $16$-MMHK against state-of-the-art Hubbard-model results for the ground-state energy and double occupancy at half filling and two doped densities in \figdisp{fig:benchmark-density}(a,b). The reference calculations use AFQMC and DMET \cite{leblanc2015solutions,zheng2017stripe,qin2016benchmark}. The mean deviation in the ground-state energy across the displayed interaction values is below $2\%$ at each density. The corresponding double-occupancy deviation is below $2\%$ at half filling and below $5\%$ for the doped densities. 

We next test the more demanding long-range magnetic response in \figdisp{fig:benchmark-density}(c), where $m=\left[S^{zz}(\mathbf Q=(\pi,\pi))\right]^{1/2}$ at half filling. Its quantitative spread relative to AFQMC and DMET is larger than that for the energy and double occupancy, but the $16$-MMHK sequence increases monotonically with $U$ and approaches the Heisenberg limit expected as $U/t\to\infty$ \cite{manousakis1991heisenberg,sandvik2026highprecision}. In contrast, the displayed AFQMC sequence is non-monotonic and the DMET value at $U/t=8$ lies above that limit, highlighting magnetic order as a more sensitive test of $16$-MMHK. Independent spinless $t$--$V$ tests of the added nearest-neighbor interaction---mixing-resolution convergence in one dimension and comparison with continuous-time-QMC and DMRG in two dimensions---are presented in Supplemental Material, Sec.~2 and Fig.~S1. Together with the accurate ground-state energy and double occupancy and the physically consistent magnetic trend above, these tests establish the static baseline for applying $V$-extended $16$-MMHK to the real-frequency GZ problem.

\subsection{Spin and charge backgrounds predict distinct GZ coefficients}
\begin{figure}[!t]
\centering
\includegraphics[width=\columnwidth]{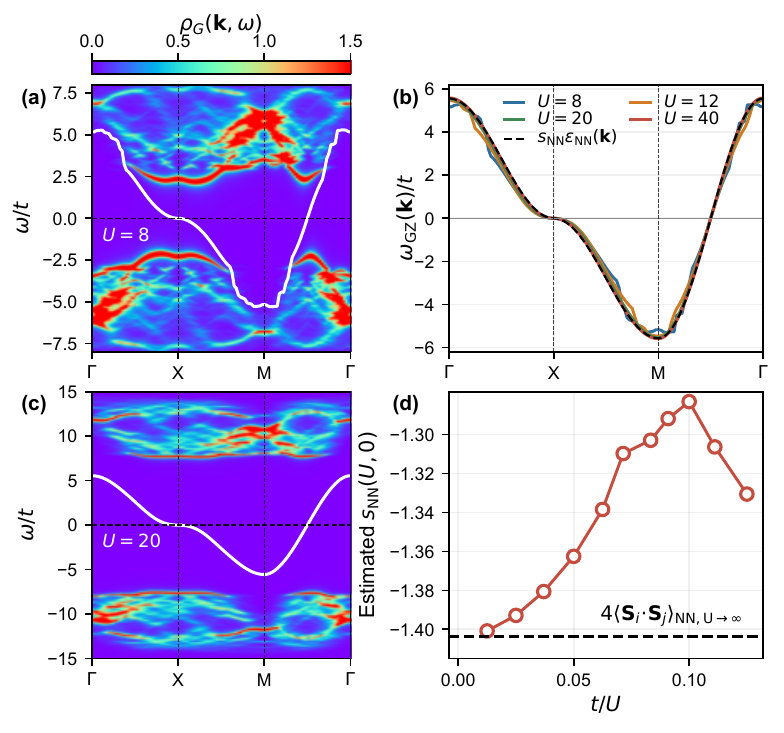}
\caption{\textbf{Mott-side GZ dispersion.} (a,c) Representative spectral functions at $V=0$ for $U=8$ and 20; white curves mark the gap-interior GZ dispersion. The frequency range expands with the interaction while the spectral color scale is fixed. (b) Extracted GZ dispersions for $U=8,20,$ and 40 and the zero-offset fit $\wzero(\mathbf{k})=s_{\mathrm{NN}}\epsk$ to the $U=40$ curve. (d) Estimated $s_{\mathrm{NN}}(U,0)$ versus $t/U$ for $U=8\sim80$. The dashed line is the independently determined prediction $4\langle\mathbf S_i\!\cdot\!\mathbf S_j\rangle_{\rm NN,U\to\infty}=-1.4036$ at the present $16$-MMHK resolution. The complete spectral sequence is shown in Supplemental Fig.~S2.}
\label{fig:mott-axis}
\end{figure}

\begin{figure}[!t]
\centering
\includegraphics[width=\columnwidth]{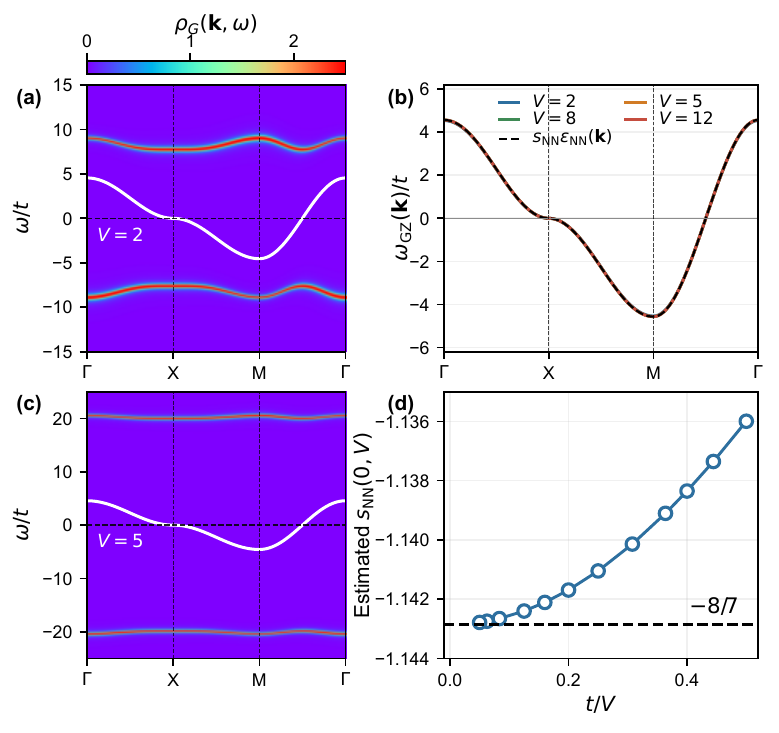}
\caption{\textbf{CDW-side GZ dispersion.} (a,c) Representative spectral functions at $U=0$ for $V=2$ and 5; white curves mark the gap-interior GZ dispersion. The frequency range expands with the interaction while the spectral color scale is fixed. (b) Extracted GZ dispersions for $V=2,5,$ and 8 and the zero-offset fit $\wzero(\mathbf{k})=s_{\mathrm{NN}}\epsk$ to the $V=8$ curve. (d) Estimated $s_{\mathrm{NN}}(0,V)$ versus $t/V$ for $V=2\sim20$. The dashed line is the parameter-free charge-defect prediction $-8/7$. The complete spectral sequence is shown in Supplemental Fig.~S3.}
\label{fig:cdw-axis}
\end{figure}

With the static benchmarks in hand, we turn to the half-filled
real-frequency Green function and its zeros computed with the
$V$-extended $16$-MMHK construction. All spectral maps use the fixed broadening $\eta=0.05$. The $U$-only data span $U=8,12,20,$ and $40$ with $V=0$; representative spectra at $U=8$ and 20 are shown in \figdisp{fig:mott-axis}(a,c), while panel (b) compares the GZ curves for all four interactions. The $V$-only data span $V=2,5,8,$ and $12$ at $U=0$; representative spectra at $V=2$ and $5$ appear in \figdisp{fig:cdw-axis}(a,c), with all four GZ curves in panel (b). The white lines mark smooth, single-valued gap-interior GZ dispersions despite the broad Hubbard continua on the Mott side and the narrowing charge-defect bands on the CDW side. We first establish why a zero must obtain and derive what fixes its leading dispersion.

In either correlated insulator, the interaction separates the single-particle spectrum into removal and addition sectors. For each resolved momentum in the present Lehmann representation, $\ReG$ is positive immediately above the highest removal pole and negative immediately below the lowest addition pole; continuity therefore enforces at least one zero in the intervening gap. Because $G^{-1}=G_0^{-1}-\Sigma$, a GZ corresponds to a self-energy pole, a nonperturbative signature of strong correlation.

At $\tp=0$, the numerically extracted GZ dispersions reveal a common
simplification. As shown for the $U$-only and $V$-only sequences in
\figdisp{fig:mott-axis}(b) and \figdisp{fig:cdw-axis}(b), respectively,
their momentum dependence is accurately described by,
\begin{equation}
  \wzero(\mathbf{k})=s_{\mathrm{NN}}\epsk,
  \label{eq:scale}
\end{equation}
where $s_{\mathrm{NN}}(U,V)$ is the signed, dimensionless, and $\mathbf{k}$-independent NN coefficient. Consequently, the zero-frequency contour $\wzero(\mathbf{k})=0$ coincides with $\epsk=0$, so the Luttinger surface is identical to the noninteracting Fermi surface at half filling, consistent with earlier work~\cite{stanescu2007luttinger}. The value and physical origin of $s_{\mathrm{NN}}$ can be determined analytically from the first spectral moments of the particle-removal and particle-addition sectors:
\begin{equation}
G(\mathbf{k},\omega)\simeq
\frac{n_{\mathbf k\sigma}}{\omega-E_h(\mathbf{k})}
+\frac{1-n_{\mathbf k\sigma}}{\omega-E_p(\mathbf{k})}.
\label{eq:two-sector}
\end{equation}
Here $E_h$ and $E_p$ are the centroids of the complete many-pole removal
and addition sectors. The zeroth-moment sum rules identify their integrated
weights as $n_{\mathbf k\sigma}$ and $1-n_{\mathbf k\sigma}$, respectively,
where $n_{\mathbf k\sigma}=\langle c^\dagger_{\mathbf k\sigma}
c_{\mathbf k\sigma}\rangle$ is the spin-resolved momentum distribution.
We do not claim that the spectrum contains only two poles; we use the
sector-centroid expansion only to locate the leading gap-interior GZ. Setting
\disp{eq:two-sector} to zero gives
\begin{equation}
\begin{aligned}
\wzero &\simeq n_{\mathbf k\sigma}E_p
 +(1-n_{\mathbf k\sigma})E_h=\mathcal C_{\mathbf{k}}+\delta_{\mathbf{k}}\Delta_{\mathbf{k}},
\end{aligned}
\label{eq:moment-relation}
\end{equation}
where $\mathcal C_{\mathbf k}=(E_p+E_h)/2$ is the midpoint of the sector
centroids relative to the chemical potential,
$\delta_{\mathbf k}=n_{\mathbf k\sigma}-1/2$ is the deviation of the
spin-resolved momentum distribution from half occupancy, and
$\Delta_{\mathbf k}=E_p-E_h$ is the centroid separation.
\Disp{eq:moment-relation} therefore separates the GZ energy into
the common sector shift and the shift produced by unequal weights across
the interaction-scale separation.
In the atomic two-pole problem, this relation is exact and coincides with
the pole in the self-energy of Ref.~\cite{dave2013absence}; here it
provides the leading term of a strong-coupling spectral-moment expansion
\cite{harris1967single,pairault2000strong}. Because
$\Delta_{\mathbf k}$ is $O(U)$ or $O(V)$ while the virtual-hopping weight
imbalance $\delta_{\mathbf k}$ is $O(t/U)$ or $O(t/V)$, their product in
\disp{eq:moment-relation} remains $O(t)$. The full moment derivation
and corrections are given in Supplemental Material, Sec.~4.

For $U\gg t$ and $V=0$, the spin-correlated Mott background in \figdisp{fig:real-space-backgrounds}(a) is predominantly singly occupied, so hopping between neighboring sites creates a virtual doublon--holon pair governed by the bond spin
correlation
$C_{\mathrm{NN}}=\langle\mathbf S_i\!\cdot\!\mathbf S_{i+\hat x}\rangle$. The Schrieffer--Wolff $t/U$ expansion \cite{macdonald1988tu} gives
$\mathcal C_{\mathbf k}
 = (1/2+2C_{\mathrm{NN}})\epsk$ and $
\delta_{\mathbf k}\Delta_{\mathbf k}
 = (2C_{\mathrm{NN}}-1/2)\epsk$.
The spin-independent $\pm\epsk/2$ terms cancel, whereas the
$2C_{\mathrm{NN}}\epsk$ terms add, yielding
\begin{equation}
\wzero^{(U)}(\mathbf{k})
 =4C_{\mathrm{NN}}\epsk+O(t^2/U).
\label{eq:spin-limit}
\end{equation}
Since $C_{\mathrm{NN}}<0$, \disp{eq:spin-limit} predicts the signed
coefficient $s_{\mathrm{NN}}=4C_{\mathrm{NN}}<0$ without fitting the
GZ curve. Details are given in Supplemental Material, Sec.~5. \Figdisp{fig:mott-axis} tests \disp{eq:spin-limit} against the full
Green function. Panels (a,c) show broad Hubbard continua coexisting
with a clean gap-interior GZ, while panels (b,d) show its NN momentum
dependence and coefficient approaching $4C_{\mathrm{NN}}$. The
relation persists down to $U=8t$ and is therefore not confined to the
formal $U\to\infty$ limit. Additional spectral snapshots and
occupied-band views are provided in Supplemental Material, Sec.~3
and Figs.~S2 and S4.

For $V\gg t$ and $U=0$, \figdisp{fig:real-space-backgrounds}(b) shows the checkerboard background of alternating doubly occupied and empty sites. A hop across a doublon--holon bond creates a defect costing
$3(2V)+V=7V$, and therefore produces the weight imbalance $\delta_{\mathbf k}=n_{\mathbf k\sigma}-1/2=-\epsk/(7V)$. There is no accompanying spin-correlation factor: on a doublon--empty bond, the allowed-hop projector is exactly $n_{i\sigma}(1-n_{j\sigma})=1$, and both atomic sites carry zero spin. The particle-addition and particle-removal moments are separated by $8V$.

Defining the signed first moment as $\mathcal M_1=(1-n_{\mathbf k\sigma})E_p+n_{\mathbf k\sigma}E_h$ (the complementary weighting to \disp{eq:moment-relation}), the exact extended-Hubbard commutator gives, to leading order in $t/V$,
$\mathcal M_1(\mathbf k)=\epsk-V\delta_{\mathbf k}+O(t^2/V)$. Together with
$\mathcal M_1=\mathcal C_{\mathbf k}-\delta_{\mathbf k}\Delta_{\mathbf k}$,
this gives
$\mathcal C_{\mathbf k}
=\epsk+7V\delta_{\mathbf k}
=O(t^2/V).$ Unlike the Mott case, here the sector center is $\mathcal C_{\mathbf k}$ dispersionless. The GZ nevertheless remains dispersive at $O(t)$ , because its leading contribution comes entirely from the spectral-weight term $\delta_{\mathbf k}\Delta_{\mathbf k}$. 
Equation~\ref{eq:moment-relation} then gives
\begin{equation}
\wzero^{(V)}(\mathbf{k})=-\frac{8}{7}\epsk+O(t^2/V).
\label{eq:charge-limit}
\end{equation}
\Disp{eq:charge-limit} fixes the parameter-free coefficient
$s_{\mathrm{NN}}(0,V\!\to\!\infty)=-8/7$ from local charge-defect
geometry. The full derivation is given in Supplemental Material,
Sec.~6.

\Figdisp{fig:cdw-axis} tests this prediction numerically. Panels (a,c)
show the particle-addition and particle-removal features narrowing
toward charge-defect bands while the gap-interior GZ remains smooth;
panels (b,d) show its NN momentum dependence and coefficient
converging to $-8/7$. Together with \figdisp{fig:mott-axis}, these
results establish two distinct strong-coupling mechanisms within the
same sector-cancellation structure of
\disp{eq:moment-relation}: spin exchange on the Mott side and local
defect kinematics on the CDW side. Additional spectra and
occupied-band views are provided in Supplemental Figs.~S3 and S5.

\begin{figure}[!t]
\centering
\includegraphics[width=0.8\columnwidth]{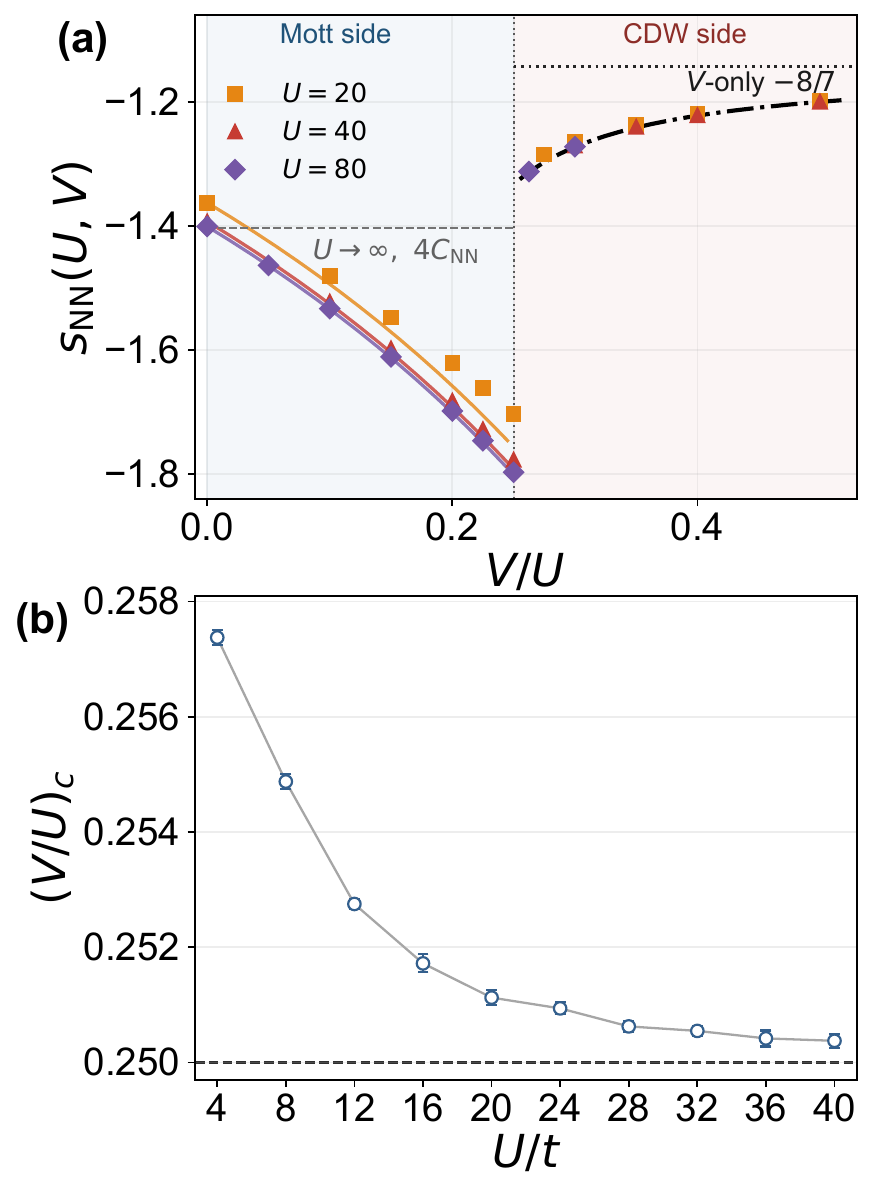}
\caption{\textbf{GZ coefficient $s_{\mathrm{NN}}$ across the Mott--CDW boundary.} (a) Signed coefficients for $U/t=20$, 40, and 80. Symbols are the GZ data. On the Mott side, the colored curves use \disp{eq:mott-law} with the independently calculated $C_{\mathrm{NN}}(U,V)$; the black dash--dotted curve on the CDW side uses the leading rational term of \disp{eq:charge-law}. Horizontal guides mark the interaction-axis limits and the vertical dotted line marks $V/U=1/4$. (b) Independently bracketed first-order transition ratio. Bars show the sampled brackets, the gray line guides the eye, and the dashed line marks $1/4$; Supplemental Material gives the ground-state diagnostics.}
\label{fig:plane}
\end{figure}

\begin{figure}[!t]
\centering
\includegraphics[width=\columnwidth]{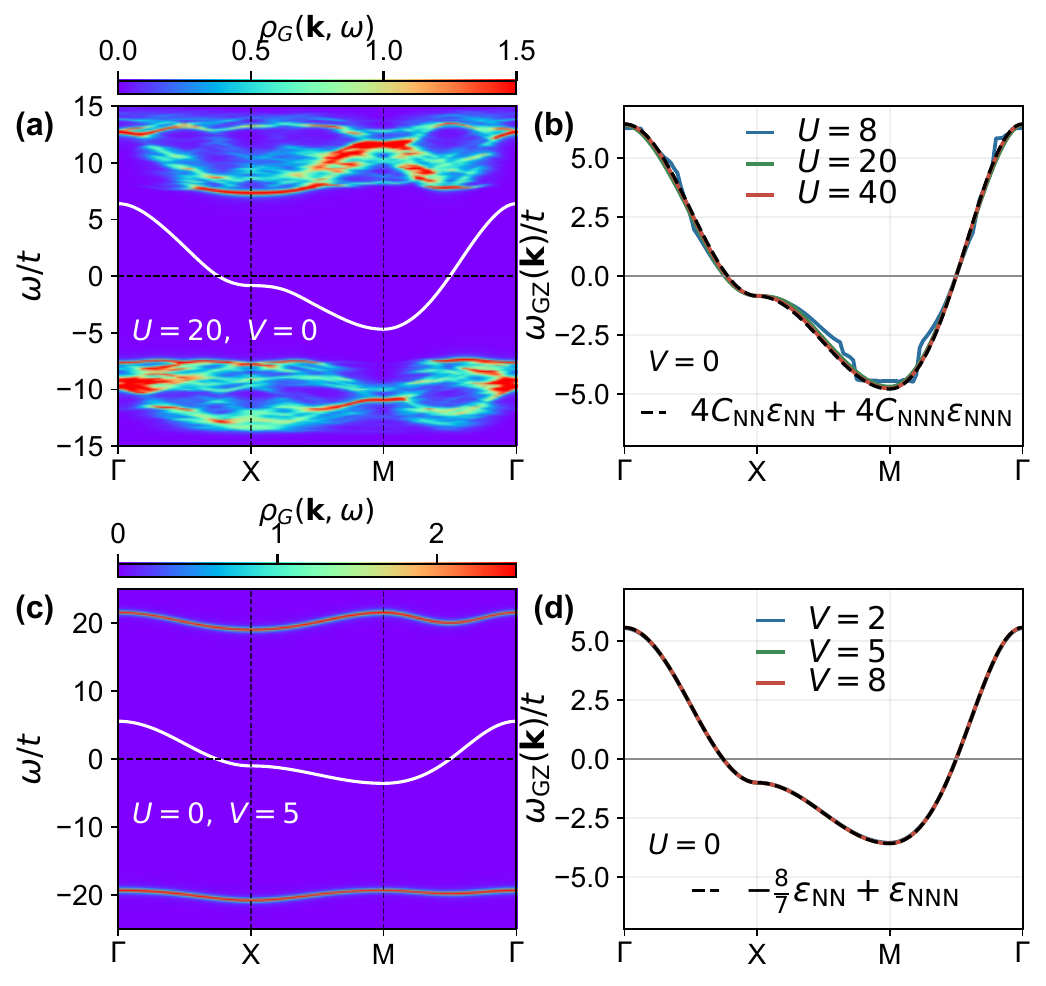}
\caption{\textbf{GZ dispersion at finite $\tp=-0.25t$.} (a) Mott-side spectrum at $U=20$, $V=0$; the white curve marks the gap-interior GZ branch. (b) Corresponding $U$-only GZ dispersions for $U=8,20,$ and 40, compared with the parameter-free $U\to\infty$ spin-correlation reference $4C_{\mathrm{NN}}\varepsilon_{\mathrm{NN}}+4C_{\mathrm{NNN}}\varepsilon_{\mathrm{NNN}}$ (dashed), with $(4C_{\mathrm{NN}},4C_{\mathrm{NNN}})=(-1.4029,0.8355)$. (c) CDW-side spectrum at $U=0$, $V=5$. (d) Corresponding $V$-only GZ dispersions for $V=2,5,$ and 8, compared with the parameter-free charge rule $-(8/7)\varepsilon_{\mathrm{NN}}+\varepsilon_{\mathrm{NNN}}$ (dashed).}
\label{fig:finite-tprime-axes}
\end{figure}

\subsection{Finite-$(U,V)$ strong-coupling laws}

\Figdisp{fig:plane}(a) shows that $|s_{\mathrm{NN}}|$ grows on the Mott side and switches discontinuously to the CDW branch as $V$ increases, although the central spectral gap stays open (Supplemental Material, Sec.~7 and Figs.~S6 and S7). This jump reflects a change in the virtual charge defect that controls the GZ dispersion. Combining the exact spectral first moment with the leading NN hopping coherence determines the strong-coupling law on each side. The notation distinguishes the neutral virtual-defect costs $\Delta E_{\mathrm M}$ and $\Delta E_{\mathrm C}$, which control the hopping coherence, from the addition--removal sector separations $\Delta_{\mathrm M}$ and $\Delta_{\mathrm C}$ entering the moment relation. On the Mott side, the virtual NN doublon--holon excitation has energy cost
$\Delta E_{\mathrm M}=U-V$. The corresponding spectral-weight imbalance is
$\delta_{\mathbf k}=(2C_{\mathrm{NN}}-1/2)\epsk/\Delta E_{\mathrm M}$, while the particle--hole separation is $\Delta_{\mathrm M}=U$. To leading order, the exact first moment becomes
$\mathcal M_1=\epsk-V\delta_{\mathbf k}+O(t^2/\Delta E_{\mathrm M})$; substituting it into
$\wzero=\mathcal M_1+2\delta_{\mathbf k}\Delta_{\mathrm M}$ gives
\begin{equation}
 s_{\mathrm{NN}}^{\mathrm M}(U,V)
 =1+\frac{2U-V}{U-V}
 \left(2C_{\mathrm{NN}}-\frac12\right)
 +O\!\left(\frac{t}{\Delta E_{\mathrm M}}\right).
 \label{eq:mott-law}
\end{equation}
The Mott-side curves in \Figdisp{fig:plane}(a) are obtained by evaluating \disp{eq:mott-law} with the independently calculated $C_{\mathrm{NN}}(U,V)$.

On the CDW side, the relevant checkerboard defect has energy cost
$\Delta E_{\mathrm C}=7V-U$, while the particle--hole separation is
$\Delta_{\mathrm C}=8V-U$. Together with
$\delta_{\mathbf k}=-\epsk/\Delta E_{\mathrm C}$, these quantities give
\begin{equation}
 s_{\mathrm{NN}}^{\mathrm C}(U,V)
 =-\frac{8V-U}{7V-U}
 +O\!\left(\frac{t}{\Delta E_{\mathrm C}}\right).
 \label{eq:charge-law}
\end{equation}
At $U=0$, the leading term becomes the parameter-free result $-8/7$. More generally, the leading term in \disp{eq:charge-law} depends only on $U/V$, so scaling $U$ and $V$ together at fixed ratio leaves the strong-coupling CDW-side coefficient unchanged. The complete derivations and numerical tests are given in Supplemental Material, Sec.~8.

The two virtual excitation energies expose the underlying physics: $V$ softens the
Mott-side doublon--holon excitation, whereas $U$ softens the CDW-side
checkerboard defect. The discontinuous change from spin-exchange physics to
checkerboard-defect kinematics therefore produces the jump in
$s_{\mathrm{NN}}$. Independent discontinuities in the charge and spin
observables and in the Hellmann--Feynman energy derivative establish the same
transition as first order (Supplemental Material, Sec.~9, Figs.~S8--S10, and
Tables~S5 and S6). \Figdisp{fig:plane}(b) shows $(V/U)_c$ decreasing from $0.2574$ at
$U/t=4$ to $0.2504$ at $U/t=40$, approaching the atomic strong-coupling value
of $1/4$.

\subsection{Frustrated hopping}
Having established the NN relation at $\tp=0$, we add next-nearest-neighbor hopping to break particle--hole symmetry and determine how the GZ dispersion generalizes. \Figdisp{fig:finite-tprime-axes}(a) shows that the resulting asymmetric Mott spectrum still contains a smooth gap-interior GZ. Writing
$\varepsilon_{\mathbf k}=\varepsilon_{\mathrm{NN}}(\mathbf k)+
\varepsilon_{\mathrm{NNN}}(\mathbf k)$, the one-coefficient relation in
\disp{eq:scale} generalizes to separate contributions from the two hopping classes. The same cancellation between the addition and removal first moments applies to each class, giving
\begin{equation}
 \wzero(\mathbf k)
 =4\sum_b C_b\varepsilon_b(\mathbf k)
 +O(t_{\max}^2/U),
 \label{eq:hopping-channel-spin-limit}
\end{equation}
where $b=\mathrm{NN},\mathrm{NNN}$,
$C_b=\langle\mathbf S_i\!\cdot\!\mathbf S_{i+\boldsymbol\delta}\rangle_{\boldsymbol\delta\in b}$,
and $t_{\max}=\max(|t|,|\tp|)$. The factor $4$ is unrelated to the coordination number: each of the two first-moment contributions produces
$2C_b\varepsilon_b$. The full derivation is given in Supplemental Material, Sec.~5.

At large $U$, the corresponding spin background is the $J_1$--$J_2$ Heisenberg model, with $J_2/J_1=(\tp/t)^2$. Because
$C_{\mathrm{NN}}<0$, the NN component is inverted. Diagonal sites belong to the same N\'eel sublattice and have
$C_{\mathrm{NNN}}>0$, so the NNN component retains the sign of
$\varepsilon_{\mathrm{NNN}}$. The numerical GZ dispersions in
\figdisp{fig:finite-tprime-axes}(b) stabilize with $U$ toward this independently calculated spin-correlation prediction. Further checks for
$\tp=-0.25t$ and $-0.4t$ are provided in Supplemental Material, Sec.~10 and Figs.~S11--S14.

The CDW-side spectrum in \figdisp{fig:finite-tprime-axes}(c) likewise retains a smooth GZ between asymmetric addition and removal bands, but its two hopping contributions have a different origin.  A NN hop creates the $7V$ defect that
produces the $-8/7$ coefficient, whereas a NNN hop propagates an
added or removed charge within one checkerboard sublattice and enters directly with unit coefficient.  Consequently,
\begin{equation}
 \wzero^{(V)}(\mathbf k)=-\frac87\varepsilon_{\mathrm{NN}}(\mathbf k)
 +\varepsilon_{\mathrm{NNN}}(\mathbf k)+O(t_{\max}^2/V).
 \label{eq:hopping-channel-charge-limit}
\end{equation}
The $V$-only sequence in \figdisp{fig:finite-tprime-axes}(d) approaches this parameter-free result; its derivation and finite-$V$ corrections are given in Supplemental Material, Sec.~6.
Thus, $\tp$ does not merely deform the GZ through a rescaled bare band. It separates contributions whose coefficients encode equal-time spin correlations on the Mott side and local charge-defect kinematics on the CDW side.

\section{Discussion}
A GZ inside an interaction gap is not merely an absence of spectral weight: its dispersion retains an $O(t)$ record of the many-body state that forms the gap. In the two strong-coupling limits, this record becomes quantitative. Mott-side spin correlations set the hopping-channel coefficients $s_b=4C_b$, whereas checkerboard defect kinematics fixes $s_{\mathrm{NN}}=-8/7$ and $s_{\mathrm{NNN}}=1$ on the CDW side. The GZ relations were obtained independently from strong-coupling expansions of the underlying extended-Hubbard Hamiltonian.  In mean-field treatments, zeros, dispersive or otherwise, are absent and hence are inadequate descriptors of Mott physics.  At leading strong-coupling order, increasing the mixing resolution therefore refines the numerical coefficients $s_b=4\langle\mathbf S_i\!\cdot\!\mathbf S_j\rangle_b$ toward their thermodynamic values without changing the momentum form of $\omega_{\rm GZ}$; the CDW-side coefficients are instead fixed by local defect geometry. Agreement between the numerical Green functions and these independently derived limits establishes a new paradigm.  Namely, dispersive zeros encode ordering tendencies, thereby completing the Mott paradigm.  Poles tell us about quasiparticles but it is through zeros that the accompanying order of Mott physics is embedded.

At finite $(U,V)$, the Mott-side magnitude grows as $V/U$ approaches $(V/U)_c\gtrsim1/4$ before switching discontinuously to the CDW-side branch. This switch coincides with discontinuities in $\nk$, kinetic energy, spin correlations, and checkerboard charge correlations that identify a first-order Mott--CDW transition for large $U$. The dynamical zero and the underlying ground state therefore reorganize together. Although both insulators lack low-energy quasiparticle poles, their GZ dispersions distinguish the spin-correlated and checkerboard state.

This mechanism survives the breaking of particle--hole symmetry. \Figdisp{fig:finite-tprime-axes} shows that the NN and NNN components of the GZ separately approach the correlation-controlled values on the Mott side and the defect-kinematic values on the CDW side. Moreover, the Mott-side relation remains quantitatively down to $U=8t$, a coupling widely used in one-band simulations of the cuprates \cite{zheng2017stripe,qin2022computational}. This robustness makes the GZ dispersion a practical organizing principle for momentum--frequency spectra: while poles have broadened into continua, the GZ supplies a clean, single-valued trajectory constrained by equal-time correlations. Since the complete single-particle spectrum determines $\ReG$ through a Kramers-Kr\"onig transform, the predicted sign-change trajectory is, in principle, accessible to real-frequency many-body solvers and sufficiently complete momentum-resolved spectroscopy. Beyond cuprates, the same correlation-to-zero correspondence offers a route to expose hidden organization in other correlated insulators. In topological Mott insulators, boundary zeros have been attributed to spinon excitations\cite{milliszeros}. Applying the present framework to boundary Green functions could directly test this attribution by determining whether the zero dispersion tracks the proposed spinon spectrum or instead obeys an equal-time correlation rule analogous to those derived here. In addition, the evolution of zeros under doping is a particularly important next problem: tracking how the GZ dispersion deforms, fragments, or reconnects with pole surfaces could reveal how a correlated insulator evolves into pseudogapped and possibly strange metallic states.  In essence, the zeros are the fingerprints of ordering in strongly correlated quantum matter.

\section{Materials and Methods}
\textbf{Ground state and Green functions.} For each $\mathbf f$-resolved momentum-mixing sector we obtain the half-filled ground state and evaluate the single-particle Green function by Lanczos exact diagonalization. Calculations use $N_\uparrow=N_\downarrow=8$ at the $4\times4$ mixing resolution. Here $Q=(0,0)$ is the total translation quantum number of the half-filled state in the 16-mode sector; particle addition and removal enter the corresponding $Q\pm q$ sectors. The checkerboard atomic background is represented by a translation-symmetric combination of its two sublattice patterns. Each particle-addition/removal continued fraction contains at most 1000 coefficients. The spectral function is $\rhoG=-2\,\mathrm{Im}(G^h+G^p)$, and the real part used for \disp{eq:zero-definition} is reconstructed from the same continued fractions \cite{gagliano1987dynamical}. Frequency windows are enlarged with interaction strength so that the central gap and both adjacent spectral features are retained.

\textbf{Data and code availability.} Machine-readable source data for every plotted curve and the analysis code underlying the figures will be made publicly available upon publication of this paper. The original MMHK implementation and source data are openly available with \refdisp{mai2026twisting}.

\textbf{Author contributions.} P.M. performed the calculations and data analysis. P.M. and P.W.P. developed the interpretation and wrote the manuscript. P.W.P. supervised the project.

\textbf{Competing interests.} The authors declare no competing interests.

\textbf{Acknowledgement} P.M. was supported by the U.S.~Department of Energy, Office of Science, Office of Basic Energy Sciences, under Award Number DE-SC0022311. P.W.P. acknowledges funding from the Research Board of the University of Illinois, CRB Award RB26125, for partial funding of this project.  We also thank Gaurav Tenkila for the benchmark work leading to Fig. 1 of the Supplement.  

\bibliographystyle{apsrev4-1}
\bibliography{references}

\end{document}


\maketitle
\tableofcontents

\section{$V$-extended MMHK construction and extended-Hubbard limit}
\label{sec:s-vextended-limit}
We show explicitly that the nearest-neighbor density interaction used in the main text is the momentum-mixing representation of the extended-Hubbard term and that the full extended-Hubbard Hamiltonian is recovered as the mixing resolution becomes complete. This extends the on-site construction of Ref.~\cite{mai2026twisting} to a finite-range density interaction.

Consider first a periodic lattice with $N$ sites and let $\nmix$ divide $N$. The $\nmix$-point mixing grid $\mathcal K_{\nmix}$ and its reduced Brillouin zone $\mathrm{rBZ}_{\nmix}$ give a unique decomposition of every lattice momentum,
\begin{equation}
 \mathbf k=\mathbf f+\mathbf K,
 \qquad
 \mathbf f\in\mathrm{rBZ}_{\nmix},
 \quad \mathbf K\in\mathcal K_{\nmix},
 \quad |\mathcal K_{\nmix}|=\nmix.
 \label{eq:s-momentum-partition}
\end{equation}
The corresponding internal-orbital transformation on the periodic mixing cell $\mathcal C_{\nmix}=\{\mathbf r_\alpha\}_{\alpha=1}^{\nmix}$ is
\begin{equation}
 \begin{aligned}
 c_{\mathbf f\alpha\sigma}
 &=\frac{1}{\sqrt{\nmix}}\sum_{\mathbf K\in\mathcal K_{\nmix}}
 e^{i\mathbf K\cdot\mathbf r_\alpha}
 c_{\mathbf f+\mathbf K,\sigma},\\
 c_{\mathbf f+\mathbf K,\sigma}
 &=\frac{1}{\sqrt{\nmix}}\sum_{\alpha=1}^{\nmix}
 e^{-i\mathbf K\cdot\mathbf r_\alpha}
 c_{\mathbf f\alpha\sigma}.
 \end{aligned}
 \label{eq:s-internal-transform}
\end{equation}
This is a unitary relabeling of all one-particle states, so the kinetic energy and chemical-potential terms are represented exactly at every $\nmix$.

For comparison, the interaction of the ordinary extended Hubbard model can be written in momentum space as
\begin{align}
 H_U={}&\frac{U}{N}\sum_{\mathbf k,\mathbf p,\mathbf q}
 c^\dagger_{\mathbf k+\mathbf q,\uparrow}c_{\mathbf k\uparrow}
 c^\dagger_{\mathbf p-\mathbf q,\downarrow}c_{\mathbf p\downarrow},
 \label{eq:s-full-hubbard-u}\\
 H_V={}&\frac{1}{2N}\sum_{\mathbf q}
 \mathcal V(\mathbf q)\,\rho(\mathbf q)\rho(-\mathbf q),
 \qquad
 \rho(\mathbf q)=\sum_{\mathbf k,\sigma}
 c^\dagger_{\mathbf k+\mathbf q,\sigma}c_{\mathbf k\sigma}.
 \label{eq:s-full-hubbard-v}
\end{align}
For a square-lattice nearest-neighbor repulsion,
\begin{equation}
 \mathcal V(\mathbf q)
 =V\sum_{\boldsymbol\delta\in\mathcal D_{\rm NN}}
 e^{i\mathbf q\cdot\boldsymbol\delta}
 =2V(\cos q_x+\cos q_y),
 \qquad
 \mathcal D_{\rm NN}=\{\pm\hat x,\pm\hat y\},
 \label{eq:s-v-form-factor}
\end{equation}
so Eq.~\ref{eq:s-full-hubbard-v} is exactly $V\sum_{\langle ij\rangle}n_i n_j$; the factor $1/2$ removes the double counting of the two directed representations of each bond.

The $\nmix$-MMHK restriction keeps all four fermion operators within one reduced-zone sector $\mathbf f$ and allows momentum transfer only by the discrete mixing vectors $\mathbf Q\in\mathcal K_{\nmix}$. With all $\mathbf K$ labels understood modulo $\mathcal K_{\nmix}$, the two interaction terms are
\begin{align}
 H_U^{(\nmix)}={}&\frac{U}{\nmix}\sum_{\mathbf f\in\mathrm{rBZ}_{\nmix}}
 \sum_{\mathbf K,\mathbf K',\mathbf Q\in\mathcal K_{\nmix}}
 c^\dagger_{\mathbf f+\mathbf K+\mathbf Q,\uparrow}
 c_{\mathbf f+\mathbf K,\uparrow}
 c^\dagger_{\mathbf f+\mathbf K'-\mathbf Q,\downarrow}
 c_{\mathbf f+\mathbf K',\downarrow},
 \label{eq:s-mmhk-u-momentum}\\
 H_V^{(\nmix)}={}&\frac{1}{2\nmix}\sum_{\mathbf f\in\mathrm{rBZ}_{\nmix}}
 \sum_{\mathbf Q\in\mathcal K_{\nmix}}
 \mathcal V(\mathbf Q)\,
 \rho_{\mathbf f}(\mathbf Q)\rho_{\mathbf f}(-\mathbf Q),
 \label{eq:s-mmhk-v-momentum}\\[-0.25em]
 \rho_{\mathbf f}(\mathbf Q)={}&
 \sum_{\mathbf K\in\mathcal K_{\nmix}}\sum_\sigma
 c^\dagger_{\mathbf f+\mathbf K+\mathbf Q,\sigma}
 c_{\mathbf f+\mathbf K,\sigma}.
 \nonumber
\end{align}
The $1/\nmix$ normalization follows from the $\nmix$-component internal Fourier transform and preserves extensivity in the sum over the $N/\nmix$ reduced-zone sectors.

Equation~\ref{eq:s-internal-transform} and the discrete orthogonality relation
\begin{equation}
 \frac{1}{\nmix}\sum_{\mathbf Q\in\mathcal K_{\nmix}}
 e^{i\mathbf Q\cdot(\mathbf r_\alpha-\mathbf r_\beta)}
 =\delta^{(\nmix)}_{\alpha\beta}
 \label{eq:s-discrete-orthogonality}
\end{equation}
make the real-space content transparent; $\delta^{(\nmix)}$ is the periodic Kronecker delta on $\mathcal C_{\nmix}$. They give
\begin{equation}
 \rho_{\mathbf f}(\mathbf Q)
 =\sum_{\alpha}e^{i\mathbf Q\cdot\mathbf r_\alpha}
 n_{\mathbf f\alpha}
 \label{eq:s-sector-density}
\end{equation}
and reduce Eqs.~\ref{eq:s-mmhk-u-momentum} and \ref{eq:s-mmhk-v-momentum} to
\begin{equation}
 H_U^{(\nmix)}+H_V^{(\nmix)}
 =\sum_{\mathbf f\in\mathrm{rBZ}_{\nmix}}
 \left[
 U\sum_\alpha n_{\mathbf f\alpha\uparrow}n_{\mathbf f\alpha\downarrow}
 +V\sum_{\langle\alpha\beta\rangle}
 n_{\mathbf f\alpha}n_{\mathbf f\beta}
 \right].
 \label{eq:s-mmhk-interaction-orbital}
\end{equation}
For the $V$ term, inserting Eq.~\ref{eq:s-v-form-factor} into Eq.~\ref{eq:s-mmhk-v-momentum} makes the $\mathbf Q$ sum enforce
$\mathbf r_\beta=\mathbf r_\alpha+\boldsymbol\delta$ modulo the periodic mixing cell. Thus the finite-$\nmix$ interaction in Eq.~\ref{eq:s-mmhk-interaction-orbital} is precisely the on-site $U$ plus nearest-neighbor $V$ term used in Eq.~(1) of the main text, independently in every $\mathbf f$ sector. The same argument applies to any finite-range density interaction after replacing $\mathcal V(\mathbf q)$ by its corresponding form factor.

The extended-Hubbard endpoint is now immediate. At $\nmix=N$, $\mathcal K_N$ is the full lattice Brillouin-zone grid, $\mathrm{rBZ}_N$ contains only one $\mathbf f$ point, and the internal label $\alpha$ runs over all physical lattice sites. Consequently,
\begin{equation}
 \boxed{
 \left.H_{\mathrm{MMHK}}^{(\nmix)}\right|_{\nmix=N}
 =\sum_{ij,\sigma}t_{ij}c^\dagger_{i\sigma}c_{j\sigma}
 +U\sum_i n_{i\uparrow}n_{i\downarrow}
 +V\sum_{\langle ij\rangle}n_i n_j
 -\mu\sum_i n_i
 \equiv H_{\mathrm{ext}} .}
 \label{eq:s-extended-hubbard-limit}
\end{equation}
The equality at the endpoint is algebraic, not an extrapolation: the restriction that all interacting momenta share one reduced-zone sector becomes vacuous once that reduced zone has collapsed to a single point. Equivalently, along a thermodynamic sequence of periodic mixing cells, $\mathcal K_{\nmix}$ becomes dense in the Brillouin zone as $\nmix\to\infty$, restoring every momentum-transfer channel of both $U$ and the short-range form factor $\mathcal V(\mathbf q)$. The $16$-MMHK calculation in the main text therefore retains thermodynamic momentum resolution through continuous $\mathbf f$ while using a finite interaction-mixing resolution on a controlled path whose complete-mixing endpoint is the extended Hubbard model.

\section{Benchmark details and spinless $t$--$V$ tests}
\subsection{Source selection for main-text Figure 2}
Main-text Figure~2 contains the complete density-resolved benchmark comparison, so it is not duplicated here. Auxiliary-field quantum Monte Carlo (AFQMC), density-matrix embedding theory (DMET), and twist-averaged boundary conditions (TABC) are abbreviated here. A single visual grammar is used throughout: color identifies density, filled circles connected by lines identify MMHK, open circles identify AFQMC or constrained-path AFQMC, and open triangles identify DMET. For constrained-path AFQMC, we first retain the largest finite TABC system at each $(n,U)$; we then compare that result with the $4\times4$ DMET energy, where available, and plot the lower energy. Double occupancy is taken from the same method and system as the selected energy whenever it was reported. If it was not, we admit only a result at exactly the same $(n,U)$. The same lowest-energy/corresponding-$D$ rule is applied to the half-filled external references. The magnetization panel remains restricted to half filling because the doped literature primarily reports stripe-resolved spin profiles, incommensurate structure factors, or broken-symmetry order parameters rather than a scalar quantity directly matching the present estimator.

\begin{table}[H]
\centering
\caption{\textbf{Benchmark sources and percentage deviations for main-text Fig.~2.} The upper block identifies every doped open marker. For each density and observable, the lower block reports the arithmetic mean of the absolute pointwise percentage deviation $100|X_{\rm MMHK}-X_{\rm ref}|/|X_{\rm ref}|$ over the common plotted interactions.}
\label{tab:benchmark-open-markers}
\small
\begin{tabular}{clcclc}
\toprule
Panel & Open marker and method & $n$ & $U/t$ & Finite calculation represented & Count \\
\midrule
a & \shortstack[l]{open circle,\\constrained-path AFQMC} & 0.875 & 6, 8, 12
  & one $20{\times}8$ TABC (stripe 10) energy per $U$ & 3 \\
\addlinespace
a & open triangle, DMET & 0.8 & 2, 4, 6, 8
  & one $4{\times}4$ embedding energy per $U$ & 4 \\
a & open triangle, DMET & 0.875 & 2, 4
  & one selected $4{\times}4$ embedding energy per $U$ & 2 \\
b & open triangle, DMET & 0.8 & 2, 4
  & $D$ corresponding to the selected DMET energy & 2 \\
b & open triangle, DMET & 0.875 & 8
  & closest same-parameter $D$ for the selected CP energy & 1 \\
\bottomrule
\end{tabular}
\medskip

\begin{tabular}{cccc}
\toprule
$n$ & Quantity & Reference points & Mean deviation (\%) \\
\midrule
1     & $E_0/(Nt)$ & 5 & 1.96 \\
1     & $D$        & 5 & 1.77 \\
1     & $m$        & 8 & 21.50 \\
0.875 & $E_0/(Nt)$ & 5 & 0.55 \\
0.875 & $D$        & 1 & 4.17 \\
0.8   & $E_0/(Nt)$ & 4 & 1.17 \\
0.8   & $D$        & 2 & 4.85 \\
\bottomrule
\end{tabular}
\par\smallskip
\begin{minipage}{0.96\linewidth}
\footnotesize
The $n=1$ magnetic mean is dominated by the weak-coupling $U/t=2$ comparison; for the six AFQMC/DMET points with $U/t\ge4$, the mean deviation is $9.34\%$. 
\end{minipage}
\end{table}

The largest system at each available $(n,U)$ is $20\times8$. Among this retained
constrained-path AFQMC energy and the $4\times4$ DMET energy, constrained-path AFQMC is lower at
$n=0.875$ and $U/t=6,8$ (and is the only retained source at $U/t=12$), whereas
DMET supplies the selected $n=0.875$ energies at $U/t=2,4$ and all four
$n=0.8$ energies. Exact corresponding DMET double occupancies exist at
$(n,U/t)=(0.8,2)$ and $(0.8,4)$. The selected constrained-path AFQMC energies do not carry double-occupancy data in the source table; the $4\times4$ DMET value at $(0.875,8)$ is therefore the sole same-parameter fallback. Error bars are the source-reported uncertainties.

\subsection{One- and two-dimensional spinless $t$--$V$ benchmarks}
We test the nearest-neighbor interaction independently of the on-site spin physics using the half-filled spinless model
\begin{equation}
 H_{tV}=-t\sum_{\langle ij\rangle}\left(c_i^\dagger c_j+\mathrm{H.c.}\right)
 +V\sum_{\langle ij\rangle}n_i n_j-\mu\sum_i n_i .
 \label{eq:s-spinless-tv}
\end{equation}
The single fermionic flavor removes the Hubbard-$U$ term while retaining the hopping and nearest-neighbor density interaction.

\begin{figure}[H]
\centering
\includegraphics[width=0.99\linewidth]{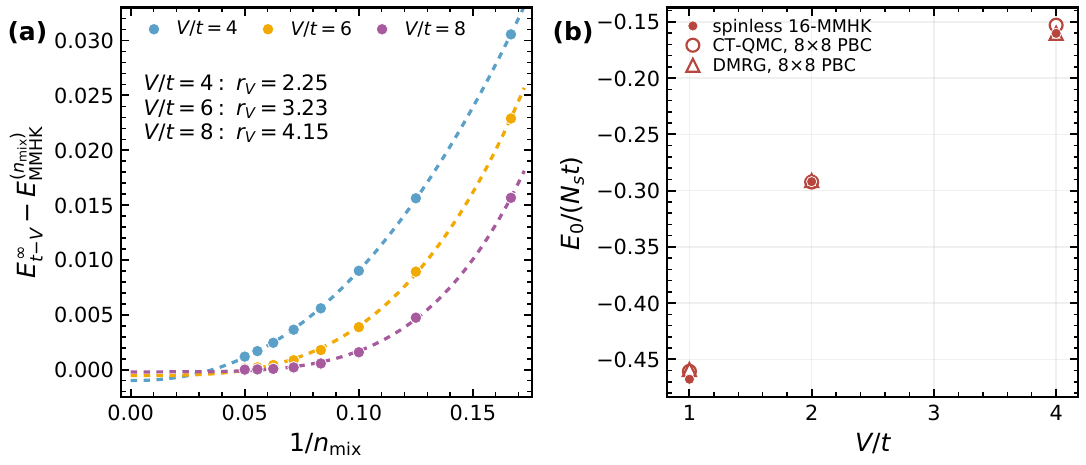}
\caption{\textbf{Spinless $t$--$V$ benchmarks of the nearest-neighbor interaction.} (a) One-dimensional mixing-resolution convergence at half filling. Symbols show $\Delta E^{(\nmix)}=E_{t-V}^{\infty}-E_{\mathrm{MMHK}}^{(\nmix)}$, where both quantities are ground-state energies per site and $E_{t-V}^{\infty}$ is the exact Bethe-ansatz limit. Dotted curves are unweighted nonlinear least-squares fits over even $\nmix=6,8,\ldots,20$ to $\Delta E^{(\nmix)}=c_V+A_V(1/\nmix)^{r_V}$. The full fitted functions are $\Delta E^{(\nmix)}=-9.840\!\times\!10^{-4}+1.776(1/\nmix)^{2.25}$ for $V/t=4$, $-5.344\!\times\!10^{-4}+7.633(1/\nmix)^{3.23}$ for $V/t=6$, and $-2\!\times\!10^{-4}+26.857(1/\nmix)^{4.15}$ for $V/t=8$. These small intercepts parameterize the displayed finite window; the exact asymptotic difference is zero. (b) Two-dimensional square-lattice ground-state energy per site at half filling for $V/t=1,2,$ and 4. Solid circles are spinless $16$-MMHK, open circles are CT-QMC, and open triangles are DMRG. The reference calculations use $N_f=32$ fermions on an $8\times8$ periodic lattice; the $16$-MMHK values use the complete 861-point symmetry-reduced mixed-momentum grid.}
\label{fig:s-spinless-tv-benchmarks}
\end{figure}

In one dimension, we use $\mu=V$ at half filling and compare the $\nmix$-MMHK ground-state energy per site with the exact thermodynamic Bethe-ansatz value \cite{yang1966xxzenergy}. The alternating odd-$\nmix$ sequence frustrates the two-sublattice charge pattern, so Fig.~\ref{fig:s-spinless-tv-benchmarks}(a) shows the unfrustrated even sequence. The displayed fits use $\nmix=6,8,\ldots,20$. Over this finite window, we allow a small intercept and fit $\Delta E^{(\nmix)}=c_V+A_V(1/\nmix)^{r_V}$, where $\Delta E^{(\nmix)}=E_{t-V}^{\infty}-E_{\mathrm{MMHK}}^{(\nmix)}$. The fitted $c_V$ remains of order $10^{-4}$, while the effective exponent $r_V$ increases with $V/t$; the exact Bethe-limit intercept remains zero. The convergence exponent is $r_V>2$ and becomes larger as $V$ increases, generally faster than the MMHK convergence to the Hubbard physics\cite{mai2026twisting}. 

We next test Eq.~\ref{eq:s-spinless-tv} on the two-dimensional square lattice with spinless $V$-extended 16-MMHK. For each $V/t=1,2,$ and 4, the energy is integrated over all 861 symmetry-reduced points of the $41\times41$ mixed-momentum grid in the reduced Brillouin zone. Figure~\ref{fig:s-spinless-tv-benchmarks}(b) compares the result with half-filled $8\times8$ periodic-boundary-condition continuous-time quantum Monte Carlo (CT-QMC) and density-matrix renormalization group (DMRG) calculations \cite{wu2024variational,wang2016stochastic}. The $16$-MMHK energy differs from CT-QMC by $1.52\%$ at $V/t=1$ and $0.12\%$ at $V/t=2$, and from DMRG by $0.06\%$ at $V/t=4$.

\section{Direct spectral comparison and occupied-band views}
\begin{figure}[H]
\centering
\includegraphics[width=\linewidth]{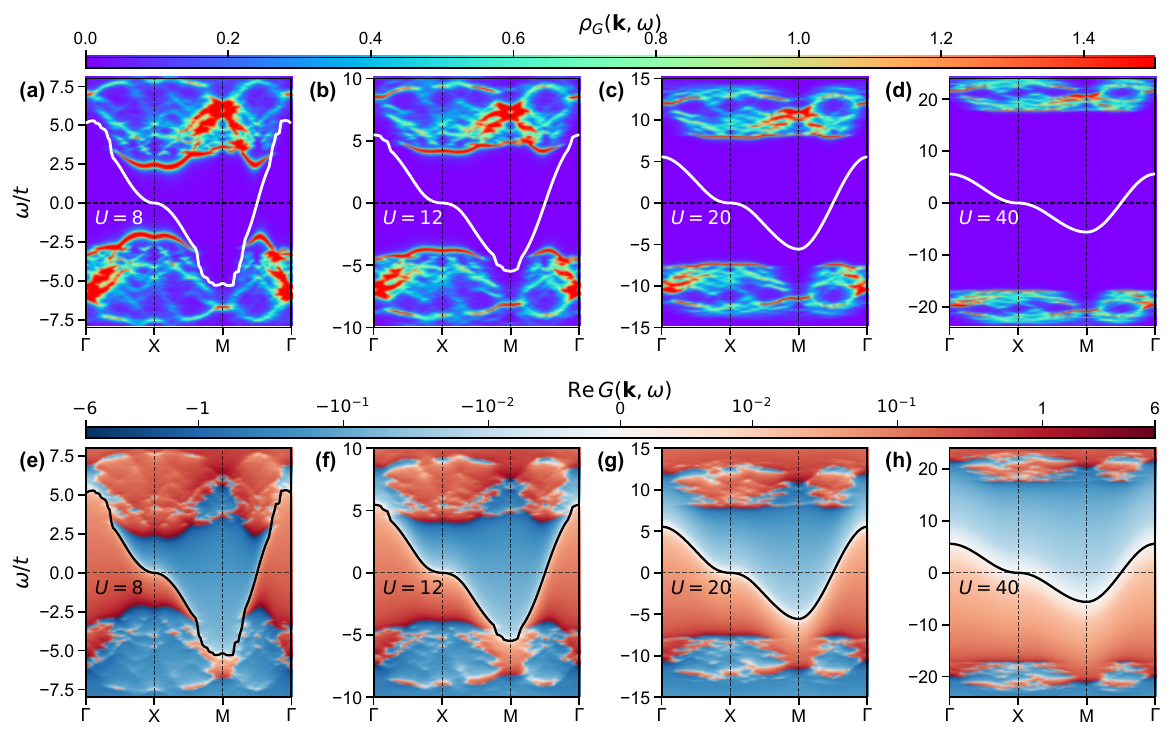}
\caption{\textbf{Complete Mott-side interaction sequence.} (a--d) Spectral functions $\rho_G(\mathbf k,\omega)$ at $V=0$ for $U/t=8,12,20,$ and 40; white curves mark the extracted gap-interior GZ dispersions. (e--h) Corresponding $\ReG(\mathbf k,\omega)$ color maps over the same frequency windows; black curves trace $\ReG=0$ and coincide with the extracted GZ branches. The frequency range expands with the interaction. A common spectral color scale is used in (a--d), and a common symmetric-logarithmic diverging scale centered at zero is used in (e--h).}
\label{fig:s-mott-axis-full}
\end{figure}

\begin{figure}[H]
\centering
\includegraphics[width=\linewidth]{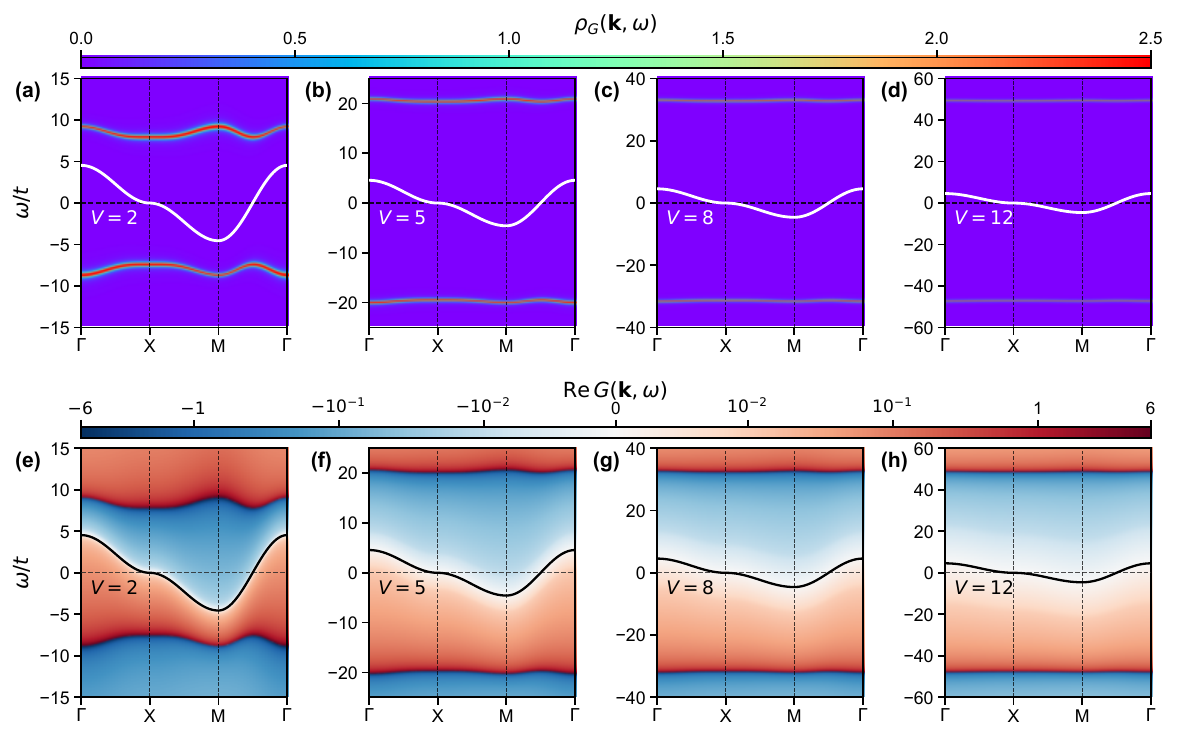}
\caption{\textbf{Complete CDW-side interaction sequence.} (a--d) Spectral functions $\rho_G(\mathbf k,\omega)$ at $U=0$ for $V/t=2,5,8,$ and 12; white curves mark the extracted gap-interior GZ dispersions. (e--h) Corresponding $\ReG(\mathbf k,\omega)$ color maps over the same frequency windows; black curves trace $\ReG=0$ and coincide with the extracted GZ branches. The frequency range expands with the interaction. A common spectral color scale is used in (a--d), and a common symmetric-logarithmic diverging scale centered at zero is used in (e--h).}
\label{fig:s-cdw-axis-full}
\end{figure}

Main-text Figs.~2 and 3 compare the two strong-coupling mechanisms through representative spectra and the complete GZ-coefficient analyses. Figures~\ref{fig:s-mott-axis-full} and \ref{fig:s-cdw-axis-full} retain the full four-interaction sequences and pair each spectral function $\rho_G(\mathbf k,\omega)$ with the corresponding $\ReG(\mathbf k,\omega)$ color map over the same frequency window. In the upper rows, white curves mark the extracted gap-interior GZ dispersions on the spectral maps; in the lower rows, the black $\ReG=0$ contours display the same branches directly. Both limits contain a gap-crossing GZ branch, but their dispersions converge to different signed coefficients of the same noninteracting NN dispersion component: $s_{\mathrm{NN}}=4C_{\mathrm{NN}}<0$ in the Mott limit and $s_{\mathrm{NN}}=-8/7$ in the CDW limit. Figures~\ref{fig:s-uonly-lower} and \ref{fig:s-vonly-lower} then isolate the occupied bands at fixed energy magnification, exposing the broad Mott continuum and narrow charge-defect bands that accompany these distinct GZ coefficients.

\subsection{On-site-interaction occupied band}
\begin{figure}[H]
\centering
\includegraphics[width=0.97\linewidth]{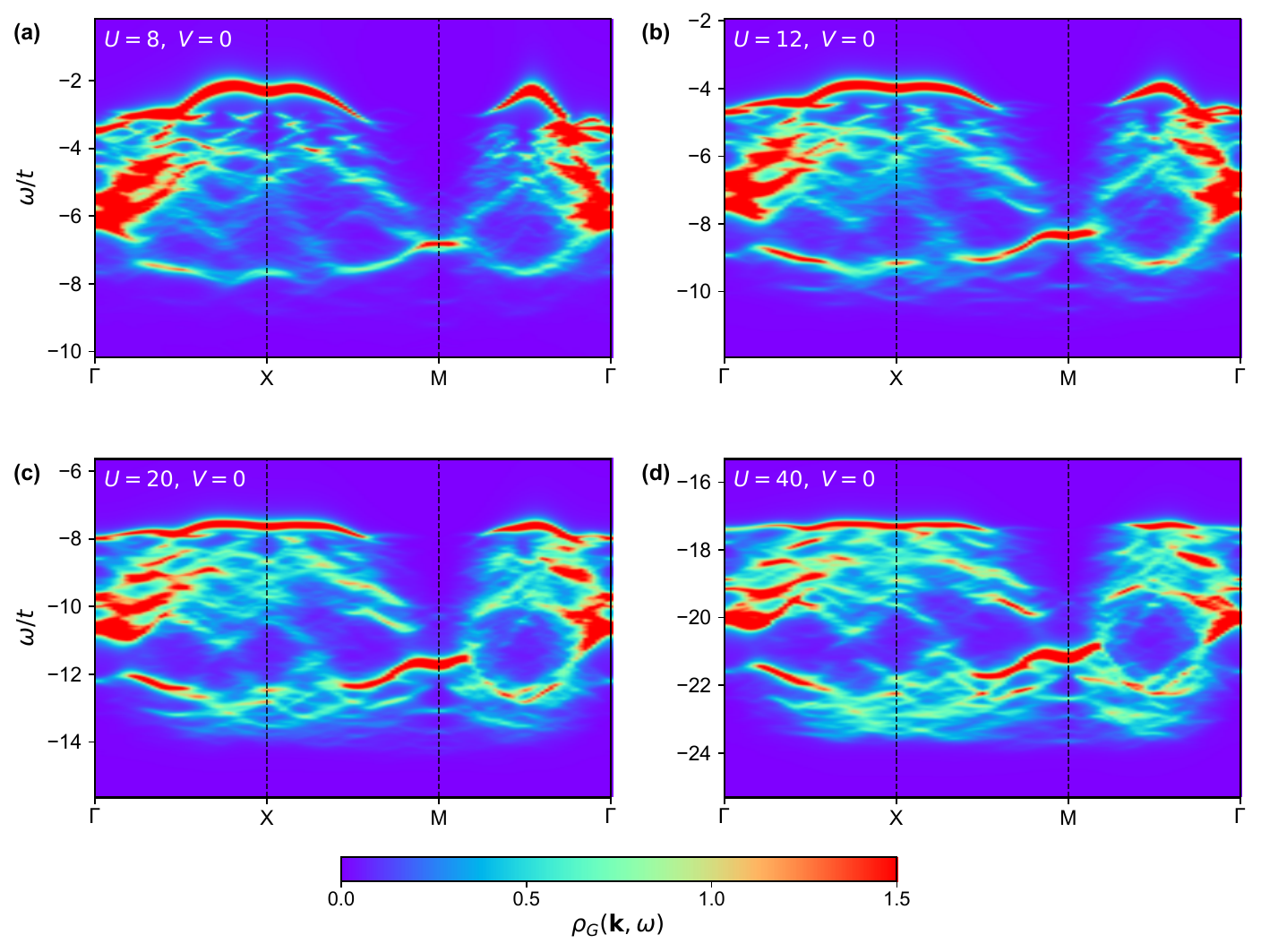}
\caption{\textbf{Mott-side occupied-band snapshots.} Lower spectral bands for $U/t=8,12,20,$ and 40 at $V=0$. Every panel has the same vertical span, $\omega_{\max}-\omega_{\min}=10t$, while its absolute window is centered on the corresponding lower-band weight. A common color normalization is used throughout.}
\label{fig:s-uonly-lower}
\end{figure}

Figure~\ref{fig:s-mott-axis-full} contains the complete on-site-interaction spectral and $\ReG$ maps together with their gap-interior GZ dispersions. Figure~\ref{fig:s-uonly-lower} instead magnifies the particle-removal band. Its four panels use a common $10t$ span, so the broad Mott-side continua are compared at fixed energy magnification while the absolute window follows the occupied spectral weight. 

\clearpage

\subsection{Nearest-neighbor-interaction occupied band}
\begin{figure}[H]
\centering
\includegraphics[width=0.97\linewidth]{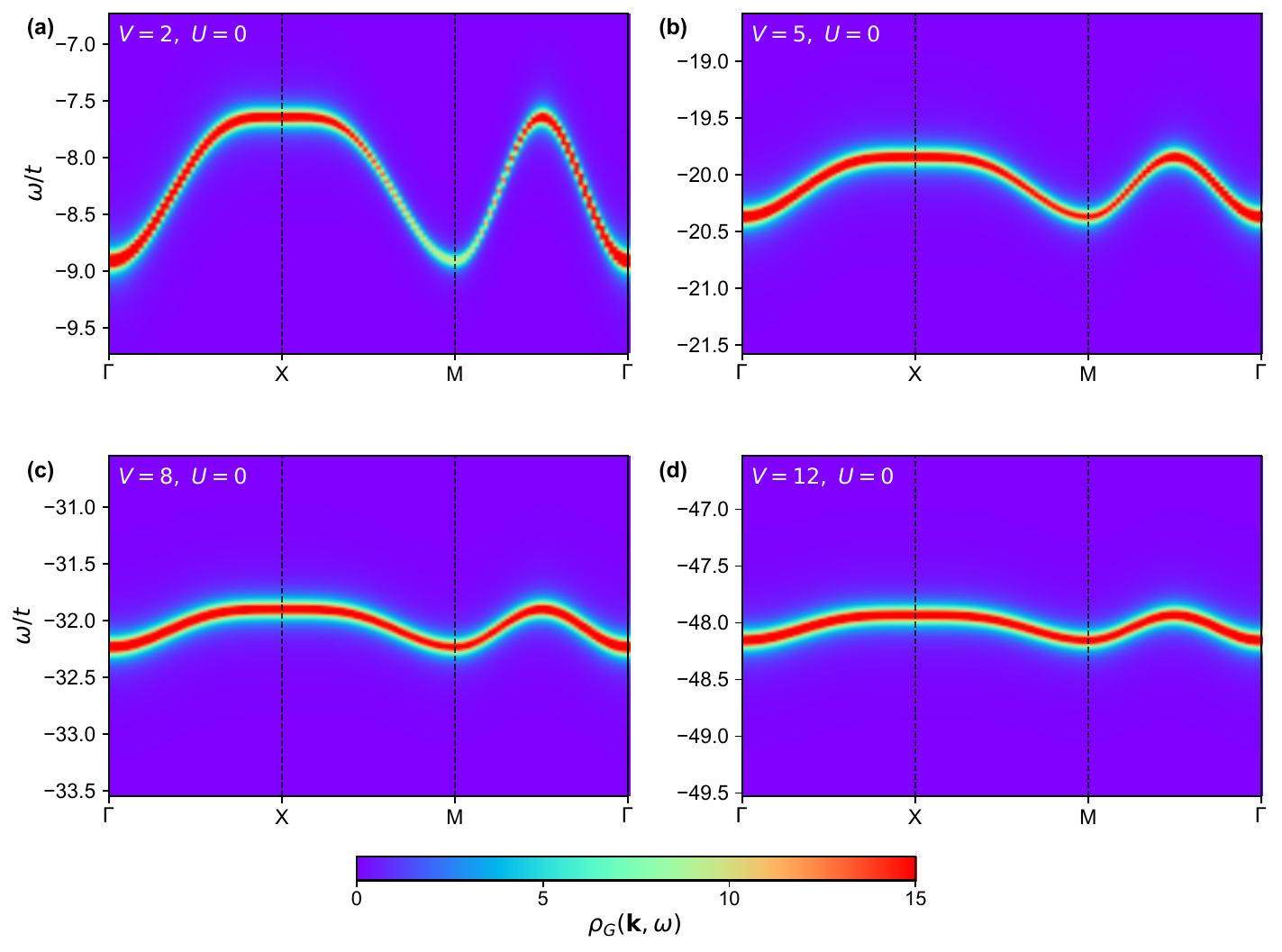}
\caption{\textbf{CDW-side occupied-band snapshots.} Lower spectral bands for $V/t=2,5,8,$ and 12 at $U=0$. Every panel has the same vertical span, $\omega_{\max}-\omega_{\min}=3t$, while the window follows the interaction-dependent removal energy. The common color scale extends to $\rho_G=15$, resolving the sharp high-intensity ridge across the sequence.}
\label{fig:s-vonly-lower}
\end{figure}

Figure~\ref{fig:s-cdw-axis-full} contains the corresponding complete nearest-neighbor-interaction spectral and $\ReG$ maps together with their GZ dispersions. The dispersive continua visible at $V/t=2$ rapidly collapse into narrow charge-addition and charge-removal bands, while the GZ remains an order-$t$ curve inside an interaction gap whose absolute size grows with $V$. Figure~\ref{fig:s-vonly-lower} magnifies the particle-removal bands for $V/t=2,5,8,$ and 12 with a common $3t$ span. The higher shared color-scale upper bound resolves the narrow high-intensity ridge instead of saturating it as in the Mott-side plot in Fig.~\ref{fig:s-uonly-lower}.

\clearpage

\section{Shared two-sector first-moment relation}
Inside either interaction gap, cancellation between particle-removal and particle-addition sectors determines the central GZ. Let $A_a(\mathbf k,E)$ be the spectral measure in sector $a=h,p$, with weight $w_a$, normalized first moment $E_a$, and central variance $\sigma_a^2$. Expanding its resolvent about $E_a$ gives
\begin{equation}
G_a(\mathbf k,\omega)=\int dE\,\frac{A_a(\mathbf k,E)}{\omega-E}
=\frac{w_a}{\omega-E_a}+\frac{w_a\sigma_a^2}{(\omega-E_a)^3}+\cdots .
\label{eq:s-resolvent-expansion}
\end{equation}
The absence of a term proportional to $(\omega-E_a)^{-2}$ follows directly from expanding about the spectral centroid rather than an arbitrary energy. Writing $\xi=E-E_a$ and using $w_a=\int dE\,A_a(\mathbf k,E)$ and $E_a=w_a^{-1}\int dE\,E A_a(\mathbf k,E)$, the first central moment is
\[
\int dE\,\xi A_a(\mathbf k,E)
=\int dE\,(E-E_a)A_a(\mathbf k,E)
=w_aE_a-E_aw_a=0.
\]
This integral is precisely the coefficient of $(\omega-E_a)^{-2}$ in the resolvent expansion. Physically, spectral weight above and below the centroid produces opposite first-order contributions that cancel. The first nonzero correction from the finite width of the sector is therefore the second central moment $w_a\sigma_a^2=\int dE\,\xi^2A_a(\mathbf k,E)$, which appears at order $(\omega-E_a)^{-3}$ in Eq.~\ref{eq:s-resolvent-expansion}. In strong coupling, the separation of particle and hole sectors is $O(U)$ or $O(V)$, while their internal widths are $O(t)$. The variance term therefore shifts a central-gap root by $O(t^2/U)$ or $O(t^2/V)$. Keeping the leading first moment in each sector gives
\begin{equation}
G(\mathbf{k},\omega)\simeq
\frac{w_h(\mathbf{k})}{\omega-E_h(\mathbf{k})}
+\frac{w_p(\mathbf{k})}{\omega-E_p(\mathbf{k})},
\label{eq:s-two-sector}
\end{equation}
where $w_h+w_p=1$ for a fixed spin. Setting the numerator of Eq.~\ref{eq:s-two-sector} to zero gives the leading root
\begin{equation}
\wzero(\mathbf{k})=w_h(\mathbf{k})E_p(\mathbf{k})+w_p(\mathbf{k})E_h(\mathbf{k}).
\label{eq:s-two-sector-root}
\end{equation}
Write
\begin{equation}
w_h=\frac12+\dk,\qquad w_p=\frac12-\dk,
\label{eq:s-sector-weights}
\end{equation}
and define
\begin{equation}
\centermom=\frac{E_p+E_h}{2},\qquad \sep=E_p-E_h.
\label{eq:s-center-separation-defs}
\end{equation}
Then
\begin{equation}
\boxed{\wzero(\mathbf{k})=\centermom+\dk\sep.}
\label{eq:s-common-zero}
\end{equation}
Here $\wzero(\mathbf k)$ is the real-frequency position of the gap-interior GZ. The sector-center term $\centermom=(E_p+E_h)/2$ is the midpoint of the complete particle-addition and particle-removal spectral centroids, measured relative to the chemical potential, and therefore records their common energy shift. The dimensionless factor $\dk=(w_h-w_p)/2$ is half the integrated spectral-weight imbalance between the removal and addition sectors, whereas $\sep=E_p-E_h$ is their centroid separation. Their product $\dk\sep$ is the displacement of the zero from the sector midpoint caused by unequal spectral weights: positive (negative) $\dk$ moves the zero toward $E_p$ ($E_h$). Although $\dk=O(t/U)$ or $O(t/V)$ in strong coupling, $\sep=O(U)$ or $O(V)$, so this weight-imbalance contribution remains $O(t)$.
Equation~\ref{eq:s-common-zero} is an algebraic identity for the two-effective-pole reconstruction and the leading relation for the full strong-coupling Green function under Eq.~\ref{eq:s-resolvent-expansion}. Its coefficients differ between the two limits because the many-body background changes the three quantities on the right-hand side. This construction follows standard strong-coupling and Lanczos moment logic \cite{harris1967single,gagliano1987dynamical,pairault2000strong}.

\subsection{Exact extended-Hubbard first moment}
\label{sec:s-finite-uv-moments}
The same notation gives an exact identity that will be used on both finite-$(U,V)$ branches and remains valid for arbitrary translationally invariant hopping, including nearest-neighbor (NN), next-nearest-neighbor (NNN), and further-neighbor terms. The products $w_pE_p$ and $w_hE_h$ define the unnormalized signed first moments of the complete particle-addition and particle-removal sectors. Their equality to the following Hamiltonian matrix elements is an exact Lehmann sum rule, not an additional approximation:
\begin{align}
 M_p=w_pE_p
 &=\langle0|c_{\mathbf k\sigma}(H-E_0)
 c^\dagger_{\mathbf k\sigma}|0\rangle,\nonumber\\
 M_h=w_hE_h
 &=-\langle0|c^\dagger_{\mathbf k\sigma}(H-E_0)
 c_{\mathbf k\sigma}|0\rangle .
 \label{eq:s-uv-signed-moments}
\end{align}
To see the particle-addition identity explicitly, let $|m,+\rangle$ be a complete set of eigenstates of the same grand-canonical Hamiltonian in the $N+1$ sector, with $H|m,+\rangle=E_m^+|m,+\rangle$. Its spectral measure is
\[
A_p(\mathbf k,E)=\sum_m
\left|\langle m,+|c^\dagger_{\mathbf k\sigma}|0\rangle\right|^2
\delta\!\left[E-(E_m^+-E_0)\right].
\]
Introduce the residue and pole energy
\[
r_m=\left|\langle m,+|c^\dagger_{\mathbf k\sigma}|0\rangle\right|^2,
\qquad \Delta_m=E_m^+-E_0.
\]
The total weight and normalized centroid are therefore
\[
w_p=\sum_m r_m,\qquad
E_p=\frac{\sum_m r_m\Delta_m}{\sum_m r_m},
\qquad
w_pE_p=\sum_m r_m\Delta_m.
\]
Thus $E_p$ is the residue-weighted average pole energy, whereas $w_pE_p$ is its unnormalized weighted sum: multiplying the normalized average by the total spectral weight exactly restores the first moment. Using completeness in the $N+1$ sector then gives
\[
\begin{aligned}
w_pE_p
&=\langle0|c_{\mathbf k\sigma}(H-E_0)
\left(\sum_m|m,+\rangle\langle m,+|\right)
c^\dagger_{\mathbf k\sigma}|0\rangle\\
&=\langle0|c_{\mathbf k\sigma}(H-E_0)
c^\dagger_{\mathbf k\sigma}|0\rangle.
\end{aligned}
\]
The projector may be omitted in the last line because $c^\dagger_{\mathbf k\sigma}|0\rangle$ lies entirely in the $N+1$ sector. Since $(H-E_0)|0\rangle=0$, the same expression can equivalently be written as $\langle0|c_{\mathbf k\sigma}[H,c^\dagger_{\mathbf k\sigma}]|0\rangle$. The removal-sector derivation is identical, except that its pole position is $E_h=-(E_n^--E_0)<0$; this produces the minus sign in the second line of Eq.~\ref{eq:s-uv-signed-moments}. Thus the equalities in Eq.~\ref{eq:s-uv-signed-moments} are exact consequences of the Lehmann representation and completeness. The fully expanded addition and removal derivation is also given in Sec.~\ref{sec:s-spin-bond-rule}, Eq.~\ref{eq:s-commutator-moments}.

Here $E_p>0$ and $E_h<0$ for the sectors bordering the central gap. Their sum satisfies
\begin{equation}
 \mathcal M_1(\mathbf k)\equiv M_p+M_h
 =w_pE_p+w_hE_h
 =\centermom-\dk\sep ,
 \label{eq:s-uv-m1-center}
\end{equation}
so Eq.~\ref{eq:s-common-zero} can equivalently be written as
\begin{equation}
 \boxed{\wzero(\mathbf k)
 =\mathcal M_1(\mathbf k)+2\dk\sep+O(t^2/\Lambda),}
 \label{eq:s-uv-root-m1}
\end{equation}
where $\Lambda$ is the relevant atomic defect energy. The remainder comes from replacing each complete spectral sector by its weight and normalized first moment, as quantified by Eq.~\ref{eq:s-resolvent-expansion}.

For the grand-canonical extended-Hubbard Hamiltonian, resolve the hopping matrix into symmetry-equivalent displacement classes $b=\mathrm{NN},\mathrm{NNN},\ldots$ and write
\[
H_t=\sum_b H_{t_b}=\sum_{ij\sigma}h_{ij}c^\dagger_{i\sigma}c_{j\sigma},
\qquad
\varepsilon(\mathbf k)=\sum_b\varepsilon_b(\mathbf k),
\]
where $\varepsilon_b(\mathbf k)$ is the Fourier transform of the hopping matrix restricted to class $b$. The exact real-space commutator is
\begin{equation}
 \boxed{
 [H,c^\dagger_{i\sigma}]
 =\sum_j h_{ji}c^\dagger_{j\sigma}
 +\left(U n_{i\bar\sigma}
 +V\sum_{\boldsymbol\delta\in\mathcal D_{\rm NN}}
 n_{i+\boldsymbol\delta}-\mu\right)c^\dagger_{i\sigma}.}
 \label{eq:s-extended-hubbard-commutator}
\end{equation}
Combining this result with the Lehmann sum rule gives
\begin{equation}
 \mathcal M_1(\mathbf k)
 =\left\langle
 \left\{c_{\mathbf k\sigma},
 [H,c^\dagger_{\mathbf k\sigma}]\right\}
 \right\rangle .
 \label{eq:s-uv-m1-commutator}
\end{equation}
The nonlocal interaction contributes a nonlocal bond-coherence term because, for $\ell\ne i$,
\begin{equation}
 \{c_{j\sigma},n_\ell c^\dagger_{i\sigma}\}
 =\delta_{ij}n_\ell-\delta_{j\ell}
 c^\dagger_{i\sigma}c_{j\sigma}.
 \label{eq:s-v-anticommutator}
\end{equation}
Define its NN form factor by
\begin{equation}
 \mathcal F_{\mathbf k}
 =\frac1{N_s}\sum_i
 \sum_{\boldsymbol\delta\in\mathcal D_{\rm NN}}
 e^{i\mathbf k\cdot\boldsymbol\delta}
 \left\langle
 c^\dagger_{i+\boldsymbol\delta,\sigma}c_{i\sigma}
 \right\rangle .
 \label{eq:s-uv-coherence-form-factor}
\end{equation}
Fourier transforming Eq.~\ref{eq:s-extended-hubbard-commutator} and using Eq.~\ref{eq:s-v-anticommutator} then yields the exact first moment
\begin{equation}
 \mathcal M_1(\mathbf k)
 =\sum_b\varepsilon_b(\mathbf k)-\mu
 +U\langle n_{i\bar\sigma}\rangle
 +V\sum_{\boldsymbol\delta}
 \langle n_{i+\boldsymbol\delta}\rangle
 -V\mathcal F_{\mathbf k}.
 \label{eq:s-uv-m1-general}
\end{equation}
At half filling on the square lattice, $\langle n_{i\bar\sigma}\rangle=1/2$ and $\langle n_i\rangle=1$, so
\begin{equation}
 \boxed{\mathcal M_1(\mathbf k)
 =\sum_b\varepsilon_b(\mathbf k)+\Delta_\mu
 -V\mathcal F_{\mathbf k},
 \qquad \Delta_\mu\equiv\frac U2+4V-\mu.}
 \label{eq:s-uv-m1-exact}
\end{equation}
This equation is exact. For particle--hole-symmetric hopping, $\mu=U/2+4V$ and $\Delta_\mu=0$, leaving $\mathcal M_1(\mathbf k)=\sum_b\varepsilon_b(\mathbf k)-V\mathcal F_{\mathbf k}$; the NN-only result is recovered by retaining only $b=\mathrm{NN}$. A particle--hole-breaking NNN or further-neighbor hopping contributes its own $\varepsilon_b(\mathbf k)$ and can also produce the momentum-independent shift $\Delta_\mu$. It does not generate a new interaction form factor because the density interaction itself remains NN, although it changes $\mathcal F_{\mathbf k}$ through the many-body state. Only in the subsequent strong-coupling expansion do we approximate $\mathcal F_{\mathbf k}$; for NN hopping, $\mathcal F_{\mathbf k}=\dk+O(t^2/\Lambda^2)$. The distinction matters because $-V\mathcal F_{\mathbf k}$ remains $O(t)$ when $U/V$ is fixed.

\section{Mott-side derivation of the hopping-channel $4\sum_b C_b\varepsilon_b$ rule}
\label{sec:s-spin-bond-rule}
At half filling and $U\gg t$, the spin background follows from a canonical
(Schrieffer--Wolff) transformation in the Hubbard $t/U$ expansion
\cite{macdonald1988tu,auerbach1994interacting}.  Write the Hubbard Hamiltonian as
\begin{equation}
\begin{aligned}
 H&=H_t+H_U-\mu N,\qquad H_U=U\hat D,
 \qquad \mu=U/2,\\
 \hat D&=\sum_i n_{i\uparrow}n_{i\downarrow},\qquad
 N=\sum_{i\sigma}n_{i\sigma}.
\end{aligned}
\label{eq:s-hubbard-doublon}
\end{equation}
and resolve the hopping according to how it changes the number of doublons,
\begin{equation}
 H_t=\mathcal T_{-1}+\mathcal T_0+\mathcal T_{+1},\qquad
 [\hat D,\mathcal T_m]=m\mathcal T_m,\qquad
 \mathcal T_{+1}^{\dagger}=\mathcal T_{-1}.
\label{eq:s-hopping-doublon-components}
\end{equation}
For an ordered-bond convention
$H_t=-\sum_{ij\sigma}t_{ij}c^\dagger_{i\sigma}c_{j\sigma}$ with
$t_{ji}=t_{ij}^*$, define
$\bar n_{i\bar\sigma}=1-n_{i\bar\sigma}$.  The three operators in
Eq.~\ref{eq:s-hopping-doublon-components} are explicitly
\begin{align}
 \mathcal T_{+1}
 &=-\sum_{ij\sigma}t_{ij}\,
 n_{i\bar\sigma}c^\dagger_{i\sigma}c_{j\sigma}
 \bar n_{j\bar\sigma},\nonumber\\
 \mathcal T_{-1}
 &=-\sum_{ij\sigma}t_{ij}\,
 \bar n_{i\bar\sigma}c^\dagger_{i\sigma}c_{j\sigma}
 n_{j\bar\sigma},\nonumber\\
 \mathcal T_0
 &=-\sum_{ij\sigma}t_{ij}\left(
 n_{i\bar\sigma}c^\dagger_{i\sigma}c_{j\sigma}n_{j\bar\sigma}
 +\bar n_{i\bar\sigma}c^\dagger_{i\sigma}c_{j\sigma}
 \bar n_{j\bar\sigma}\right).
\label{eq:s-explicit-doublon-components}
\end{align}
The ordered sum includes both hopping directions, so no additional Hermitian
conjugate is required.  In $\mathcal T_{+1}$ the arriving electron creates a
doublon while its departure creates a holon; $\mathcal T_{-1}$ is the reverse
process.  The two terms in $\mathcal T_0$ respectively move an existing
doublon or move an electron without creating one, leaving the total doublon
number unchanged.  Choose the anti-Hermitian generator
\begin{equation}
 \mathcal S=\frac{\mathcal T_{+1}-\mathcal T_{-1}}{U}
 +O(t^2/U^2).
\label{eq:s-sw-generator}
\end{equation}
Because each $\mathcal T_m$ conserves total particle number,
$[\mathcal S,N]=0$.  Together with
$[\mathcal S,H_U]=-(\mathcal T_{+1}+\mathcal T_{-1})$, this means that the transformed
Hamiltonian $\bar H=e^{\mathcal S}He^{-\mathcal S}$ has no matrix element
between sectors with different doublon number at first order in $t/U$.
Expanding the transformation through second order gives
\begin{equation}
\begin{aligned}
 \bar H
 &=H+[\mathcal S,H]+\frac12[\mathcal S,[\mathcal S,H]]+\cdots,\\
 P_0\bar H P_0
 &=-\mu N_sP_0
 -\frac1U P_0\mathcal T_{-1}\mathcal T_{+1}P_0
 +O(t^3/U^2),
\end{aligned}
\label{eq:s-sw-bch-projection}
\end{equation}
where $P_0$ projects onto the half-filled subspace with exactly one electron
per site, so $P_0NP_0=N_sP_0$.  The first term is a common grand-canonical
constant in this subspace and is omitted from the nontrivial spin Hamiltonian
below.  In obtaining the second line we used
$P_0\mathcal T_0P_0=0$, $\mathcal T_{-1}P_0=0$, and
\begin{equation}
\begin{aligned}
 \frac12[\mathcal S,\mathcal T_{+1}+\mathcal T_{-1}]
 &=\frac{1}{2U}
 [\mathcal T_{+1}-\mathcal T_{-1},
  \mathcal T_{+1}+\mathcal T_{-1}]
 +O(t^3/U^2)\\
 &=\frac{1}{2U}\Bigl(
 [\mathcal T_{+1},\mathcal T_{+1}]
 +[\mathcal T_{+1},\mathcal T_{-1}]
 -[\mathcal T_{-1},\mathcal T_{+1}]
 -[\mathcal T_{-1},\mathcal T_{-1}]\Bigr)
 +O(t^3/U^2)\\
 &=\frac1U[\mathcal T_{+1},\mathcal T_{-1}]
 +O(t^3/U^2),\\[2pt]
 P_0[\mathcal T_{+1},\mathcal T_{-1}]P_0
 &=P_0\mathcal T_{+1}\mathcal T_{-1}P_0
  -P_0\mathcal T_{-1}\mathcal T_{+1}P_0\\
 &=-P_0\mathcal T_{-1}\mathcal T_{+1}P_0.
\end{aligned}
\label{eq:s-sw-commutator}
\end{equation}
In the last line, the first product vanishes because
$\mathcal T_{-1}P_0=0$.  This gives the projected second-order term in
Eq.~\ref{eq:s-sw-bch-projection} directly.

The remaining operator has a transparent two-site action.  On a bond
$\langle ij\rangle$, a triplet is Pauli blocked, whereas a singlet couples to
the two intermediate states with a doublon on $i$ or on $j$.  Equivalently,
within the singly occupied subspace,
\begin{equation}
 P_0\mathcal T_{-1}^{(ij)}\mathcal T_{+1}^{(ij)}P_0
 =4t^2\left(\frac14-\mathbf S_i\cdot\mathbf S_j\right),
\label{eq:s-singlet-projector}
\end{equation}
because $1/4-\mathbf S_i\cdot\mathbf S_j$ is the singlet projector.  Thus
\begin{equation}
 H_{\rm spin}=\frac{4t^2}{U}\sum_{\langle ij\rangle}
 \left(\mathbf S_i\cdot\mathbf S_j-\frac14\right)
 +O(t^4/U^3).
\label{eq:s-heisenberg-effective}
\end{equation}
The Schrieffer--Wolff block diagonalization itself is not restricted to half filling; it requires that the charge sectors being eliminated remain separated from the chosen low-energy manifold by energies large compared with the hopping. Half filling is special because the no-doublon manifold contains exactly one electron per site. Consequently $P_0\mathcal T_0P_0=0$, there is no projected hopping at order $t$, and the low-energy Hamiltonian reduces to the spin model in Eq.~\ref{eq:s-heisenberg-effective}. Away from half filling, the no-doublon manifold also contains mobile holes and $P_0\mathcal T_0P_0\ne0$. For hole doping the same transformation instead gives the $t$--$J$ structure
\[
H_{\rm eff}=P_0\mathcal T_0P_0
+\sum_{ij}J_{ij}\left(\mathbf S_i\!\cdot\!\mathbf S_j
-\frac14n_in_j\right)
+H_{\rm 3site}+O(t^3/U^2),
\qquad J_{ij}=\frac{4|t_{ij}|^2}{U},
\]
where $H_{\rm 3site}=O(t^2/U)$ denotes correlated three-site hopping. Electron doping can be treated by a particle--hole transformation or by projecting onto the appropriate fixed-minimum-doublon manifold. For an extended-Hubbard model, the relevant denominators contain both $U$ and the density interactions; all charge-defect energies associated with the eliminated configurations must remain large compared with the hopping. Thus $U\gg t$ alone does not justify a single-manifold expansion when competing charge backgrounds become nearly degenerate. The spin-only reduction used below, and hence the present correlation-only GZ coefficient, is specific to the half-filled Mott manifold even though the Schrieffer--Wolff method is more general.

Odd-order terms vanish for the present half-filled nearest-neighbor bipartite
model, so the leading correction to this Heisenberg Hamiltonian is fourth
order.  The many-body quantity entering the Green-function moments is
therefore the nearest-neighbor correlator of this spin background.  We denote
it by
\begin{equation}
\Cnn=\langle\mathbf S_i\cdot\mathbf S_{i+\hat x}\rangle.
\label{eq:s-cnn}
\end{equation}
Starting from the spin correlator in Eq.~\ref{eq:s-cnn}, the corresponding
nearest-neighbor one-body density matrix follows by introducing a separate
hopping parameter $t_{ij}$ on one bond $\langle ij\rangle$,
\begin{equation}
 H^{\rm kin}_{ij}=-t_{ij}T_{ij},\qquad
 T_{ij}=\sum_\sigma
 \left(c^\dagger_{i\sigma}c_{j\sigma}
 +c^\dagger_{j\sigma}c_{i\sigma}\right).
\label{eq:s-bond-kinetic-def}
\end{equation}
This bond parameter is only a differentiation device and is set to the uniform
value $t_{ij}=t$ at the end.  Eliminating virtual doublon--holon states to
second order gives, within the singly occupied subspace,
\begin{equation}
 H^{\rm eff}_{ij}=J_{ij}\left(\mathbf S_i\cdot\mathbf S_j-\frac14\right),
 \qquad J_{ij}=\frac{4t_{ij}^2}{U}.
\label{eq:s-bond-effective}
\end{equation}
The subtraction of $1/4$ is essential: a virtual hop is allowed for an
antiparallel-spin bond but is Pauli blocked for a parallel-spin bond.

For a normalized eigenstate, the Hellmann--Feynman theorem states
\begin{equation}
 \frac{\partial E_0}{\partial t_{ij}}
 =\left\langle\Psi_0\left|
 \frac{\partial H}{\partial t_{ij}}
 \right|\Psi_0\right\rangle .
\label{eq:s-hellmann-feynman}
\end{equation}
To see the cancellation explicitly, differentiate
$E_0=\langle\Psi_0|H|\Psi_0\rangle$ and use
$H|\Psi_0\rangle=E_0|\Psi_0\rangle$:
\begin{equation}
\begin{aligned}
 \frac{\partial E_0}{\partial t_{ij}}
 ={}&\langle\partial_{t_{ij}}\Psi_0|H|\Psi_0\rangle
 +\left\langle\Psi_0\left|\frac{\partial H}{\partial t_{ij}}\right|\Psi_0\right\rangle
 +\langle\Psi_0|H|\partial_{t_{ij}}\Psi_0\rangle\\
 ={}&E_0\,\frac{\partial}{\partial t_{ij}}
 \langle\Psi_0|\Psi_0\rangle
 +\left\langle\Psi_0\left|\frac{\partial H}{\partial t_{ij}}\right|\Psi_0\right\rangle
 =\left\langle\frac{\partial H}{\partial t_{ij}}\right\rangle .
\end{aligned}
\label{eq:s-hellmann-feynman-proof}
\end{equation}
Thus the derivative acts only on the explicit $t_{ij}$ dependence of the
Hamiltonian; in particular, no additional term proportional to
$\partial\Cnn/\partial t_{ij}$ appears. Applying the theorem first to the
original Hubbard Hamiltonian and then to its low-energy spin Hamiltonian gives
two evaluations of the same energy derivative,
\begin{equation}
\begin{aligned}
 \frac{\partial E_0}{\partial t_{ij}}
 &=-\langle T_{ij}\rangle,\\
 \frac{\partial E_0^{\rm eff}}{\partial t_{ij}}
 &=\frac{\partial J_{ij}}{\partial t_{ij}}
 \left\langle\mathbf S_i\cdot\mathbf S_j-\frac14\right\rangle
 =\frac{8t}{U}\left(\Cnn-\frac14\right)
 +O(t^3/U^3).
\end{aligned}
\label{eq:s-energy-derivatives}
\end{equation}
Here the first line follows from
$\partial H^{\rm kin}_{ij}/\partial t_{ij}=-T_{ij}$, whereas the second uses
$\partial J_{ij}/\partial t_{ij}=8t_{ij}/U$ before setting $t_{ij}=t$.
The effective and Hubbard ground-state energies have the same $O(t^2/U)$
term; the omitted half-filled bipartite contribution starts at
$O(t^4/U^3)$, so the corresponding derivative error is $O(t^3/U^3)$.
Equating the two evaluations therefore gives
\begin{equation}
 -\sum_\sigma\left\langle
 c^\dagger_{i\sigma}c_{j\sigma}+c^\dagger_{j\sigma}c_{i\sigma}
 \right\rangle
 =\frac{8t}{U}\left(\Cnn-\frac14\right)+O(t^3/U^3).
\label{eq:s-bond-hf-identity}
\end{equation}
For a time-reversal-invariant, spin-symmetric state, the two spin components
are equal and each directed correlator is real.  Thus the four terms on the
left are all
$g_{ij}=\langle c^\dagger_{i\sigma}c_{j\sigma}\rangle$, yielding
$-4g_{ij}=(8t/U)(\Cnn-1/4)+O(t^3/U^3)$.  By square-lattice
symmetry, define the nearest-neighbor bond coherence as
\begin{equation}
\begin{aligned}
g_{\mathrm{NN}}
&\equiv\langle c^\dagger_{i\sigma}c_{i+\hat x,\sigma}\rangle
=\langle c^\dagger_{i\sigma}c_{i+\hat y,\sigma}\rangle\\
&=\frac{2t}{U}\left(\frac14-\Cnn\right)+O(t^3/U^3).
\end{aligned}
\label{eq:s-density-matrix-u}
\end{equation}
The signs provide a useful check: an antiferromagnetic bond has
$\Cnn<1/4$, hence $g_{\mathrm{NN}}>0$ for $t>0$, and its kinetic energy
$-t\langle T_{ij}\rangle=-4tg_{\mathrm{NN}}$ is negative.  This completes the
Hellmann--Feynman step from Eq.~\ref{eq:s-cnn} to
Eq.~\ref{eq:s-density-matrix-u}.
The absence of an $O(t^2/U^2)$ term follows from the bipartite
particle--hole symmetry of the half-filled nearest-neighbor model: the bond
coherence is odd under $t\to-t$.

For completeness, the Fourier transform can be written explicitly as
\begin{equation}
\begin{aligned}
n_{\mathbf k\sigma}
&=\frac{1}{N_s}\sum_{ij}e^{i\mathbf k\cdot(\mathbf r_i-\mathbf r_j)}
  \langle c^\dagger_{i\sigma}c_{j\sigma}\rangle\\
&=\frac12+2g_{\mathrm{NN}}(\cos k_x+\cos k_y)+O(t^3/U^3).
\end{aligned}
\label{eq:s-nk-fourier}
\end{equation}
The first line of Eq.~\ref{eq:s-nk-fourier} contains every separation
$i-j$; retaining only the NN term in the second line is a controlled
strong-coupling reduction.  Let $|\Phi_0\rangle$ lie in the half-filled
no-doublon manifold.  To first order,
\[
 |\Psi_0\rangle=|\Phi_0\rangle
 -\frac{1}{U}\mathcal T_{+1}|\Phi_0\rangle+O(t^2/U^2).
\]
Because $\mathcal T_{+1}$ contains only NN hops, it creates a doublon--holon
pair only on an NN bond.  The first-order matrix element
\[
 -\frac{2}{U}\,\operatorname{Re}
 \langle\Phi_0|c^\dagger_{i\sigma}c_{j\sigma}
 \mathcal T_{+1}|\Phi_0\rangle
\]
is therefore nonzero only when $i$ and $j$ are the same NN bond: the
one-body operator must reverse the virtual hop and return the state to the
no-doublon manifold.  Hence a non-NN coherence has no $O(t/U)$ term.

Particle--hole symmetry removes the apparent $O(t^2/U^2)$ possibility as
well.  Under
$c_{i\sigma}\mapsto\eta_i c^\dagger_{i\sigma}$, where
$\eta_i=+1$ on sublattice $A$ and $-1$ on sublattice $B$, an off-site
coherence obeys
\[
 g_{ij}\equiv\langle c^\dagger_{i\sigma}c_{j\sigma}\rangle
 =-\eta_i\eta_j g_{ji}.
\]
Time-reversal and inversion symmetry give $g_{ji}=g_{ij}\in\mathbb R$, so
$g_{ij}=0$ exactly for sites on the same sublattice.  In particular, the
diagonal $(1,1)$ and axial $(2,0)$ coherences vanish.  The nearest
symmetry-allowed non-NN separations, such as $(2,1)$ and $(3,0)$, connect
opposite sublattices and require at least three successive NN virtual hops;
they begin at $O(t^3/U^3)$.  Corrections to $g_{\mathrm{NN}}$ are likewise
odd in $t$ and next occur at this order.  This proves the remainder shown in
Eq.~\ref{eq:s-nk-fourier}, rather than assuming that all longer-range
coherences vanish.

As a direct numerical check, inverse Fourier transformation of the
all-momentum $U=40$, $V=0$ data gives, per spin,
$g_{\mathrm{NN}}\simeq2.970\times10^{-2}$,
$g_{(2,1)}\simeq-2.000\times10^{-4}$, and
$g_{(3,0)}\simeq-1.263\times10^{-4}$; the two distance-three coherences are
only $0.67\%$ and $0.43\%$ of $|g_{\mathrm{NN}}|$, respectively, while the
same-sublattice coherences vanish to numerical precision.  Thus the four NN
displacements produce the complete leading term
$2g_{\mathrm{NN}}(\cos k_x+\cos k_y)$.  Using
$\epsk=-2t(\cos k_x+\cos k_y)$ together with
Eq.~\ref{eq:s-density-matrix-u} therefore yields
\begin{equation}
n_{\mathbf{k}\sigma}=\frac12+\frac{2\Cnn-1/2}{U}\epsk+O(t^3/U^3).
\label{eq:s-nk-u}
\end{equation}
The zeroth-moment Lehmann sums identify the total removal and addition
weights with the occupied and unoccupied probabilities, respectively.  Using
Eq.~\ref{eq:s-sector-weights}, the weight imbalance is therefore
\begin{equation}
\begin{aligned}
w_h
&=\sum_n\left|\langle n,-|c_{\mathbf k\sigma}|0\rangle\right|^2
=\langle c^\dagger_{\mathbf k\sigma}c_{\mathbf k\sigma}\rangle
=n_{\mathbf k\sigma},\\
w_p
&=\sum_m\left|\langle m,+|c^\dagger_{\mathbf k\sigma}|0\rangle\right|^2
=\langle c_{\mathbf k\sigma}c^\dagger_{\mathbf k\sigma}\rangle
=1-n_{\mathbf k\sigma},\\
\dk
&=\frac{w_h-w_p}{2}=n_{\mathbf k\sigma}-\frac12\\
&=\frac{2\Cnn-1/2}{U}\epsk+O(t^3/U^3).
\end{aligned}
\label{eq:s-delta-u}
\end{equation}
The addition--removal separation is $\sep=U+O(t)$, and therefore
\begin{equation}
\dk\sep=(2\Cnn-1/2)\epsk+O(t^2/U).
\label{eq:s-weight-u}
\end{equation}
The normalized first moments can be evaluated without resolving individual
poles.  Throughout this work $H$ is the grand-canonical Hamiltonian; in the
present $V=0$ section it is $H=H_{\rm kin}+H_U-\mu N$ with $\mu=U/2$.
Let $H|0\rangle=E_0|0\rangle$, and let $|m,+\rangle$ and $|n,-\rangle$ be
complete eigenstates of the same $H$ in the $N+1$ and $N-1$ sectors, with
energies $E_m^+$ and $E_n^-$.  Start from the exact zero-temperature
retarded Green function and insert those two complete sets:
\begin{equation}
\begin{aligned}
G(\mathbf k,\omega)
={}&\left\langle0\left|c_{\mathbf k\sigma}
 \frac{1}{\omega-(H-E_0)+i0^+}
 c^\dagger_{\mathbf k\sigma}\right|0\right\rangle+\left\langle0\left|c^\dagger_{\mathbf k\sigma}
 \frac{1}{\omega+(H-E_0)+i0^+}
 c_{\mathbf k\sigma}\right|0\right\rangle\\
={}&\sum_m
\frac{\left|\langle m,+|c^\dagger_{\mathbf k\sigma}|0\rangle\right|^2}
{\omega-(E_m^+-E_0)+i0^+}+\sum_n
\frac{\left|\langle n,-|c_{\mathbf k\sigma}|0\rangle\right|^2}
{\omega+(E_n^--E_0)+i0^+}.
\end{aligned}
\end{equation}
The first sum has particle-addition poles at $E_m^+-E_0>0$; the second has
particle-removal poles at $-(E_n^--E_0)<0$.  Multiplying each pole position
by its residue and summing gives the signed first moments.  Thus, with
$M_p=w_pE_p$ and $M_h=w_hE_h$, completeness gives
\begin{equation}
\begin{aligned}
M_p=w_pE_p
&=\sum_m(E_m^+-E_0)
 \left|\langle m,+|c^\dagger_{\mathbf k\sigma}|0\rangle\right|^2\\
&=\langle0|c_{\mathbf k\sigma}(H-E_0)
 c^\dagger_{\mathbf k\sigma}|0\rangle
=\langle0|c_{\mathbf k\sigma}[H,c^\dagger_{\mathbf k\sigma}]|0\rangle,\\
M_h=w_hE_h
&=-\sum_n(E_n^--E_0)
 \left|\langle n,-|c_{\mathbf k\sigma}|0\rangle\right|^2\\
&=-\langle0|c^\dagger_{\mathbf k\sigma}(H-E_0)
 c_{\mathbf k\sigma}|0\rangle
=-\langle0|c^\dagger_{\mathbf k\sigma}[H,c_{\mathbf k\sigma}]|0\rangle.
\end{aligned}
\label{eq:s-commutator-moments}
\end{equation}
Equation~\ref{eq:s-commutator-moments} is the exact first spectral-moment
sum rule following from the Lehmann representation.  It is used because it
determines the particle and hole moment centers from equal-time commutators;
no individual pole needs to be identified.  The strong-coupling approximation
enters only later, when the full spectral measure of each sector is represented
by its weight and normalized first moment.

\paragraph{From Eq.~\ref{eq:s-commutator-moments} to the moment center.}
For this grand-canonical Hubbard Hamiltonian,
\begin{equation}
\begin{aligned}
\bigl[H,c^\dagger_{\mathbf k\sigma}\bigr]
={}&(\epsk-\mu)c^\dagger_{\mathbf k\sigma}
+\frac{U}{\sqrt{N_s}}\sum_i e^{i\mathbf k\cdot\mathbf r_i}
n_{i\bar\sigma}c^\dagger_{i\sigma},\\
\bigl[H,c_{\mathbf k\sigma}\bigr]
={}&-(\epsk-\mu)c_{\mathbf k\sigma}
-\frac{U}{\sqrt{N_s}}\sum_i e^{-i\mathbf k\cdot\mathbf r_i}
n_{i\bar\sigma}c_{i\sigma}.
\end{aligned}
\label{eq:s-hubbard-commutators}
\end{equation}
Adding the two weighted moments in Eq.~\ref{eq:s-commutator-moments} produces the exact first-moment sum rule
\begin{equation}
M_p+M_h
=\left\langle
\left\{c_{\mathbf k\sigma},
[H,c^\dagger_{\mathbf k\sigma}]
\right\}\right\rangle
=\epsk-\mu+U\langle n_{i\bar\sigma}\rangle
=\epsk ,
\label{eq:s-first-moment-sum}
\end{equation}
where the last equality uses $\mu=U/2$ and
$\langle n_{i\bar\sigma}\rangle=1/2$. At $t=0$, the two weighted moments are
$M_p=U/4$ and $M_h=-U/4$. For nearest-neighbor hopping on the bipartite
lattice, particle--hole symmetry maps particle addition at $\mathbf k$ to
particle removal at $\mathbf k+\mathbf Q$, where
$\mathbf Q=(\pi,\pi)$ and $\varepsilon_{\mathrm{NN}}(\mathbf k+\mathbf Q)=-\epsk$.
Consequently, the $O(t)$ part of the sum rule is divided equally between
the two moments:
\begin{equation}
M_p=\frac{U}{4}+\frac{\epsk}{2}+O(t^2/U),
\qquad
M_h=-\frac{U}{4}+\frac{\epsk}{2}+O(t^2/U).
\label{eq:s-weighted-moments-u}
\end{equation}

To expose how the spin correlation enters after normalization, define
$d=2\Cnn-1/2$. Equation~\ref{eq:s-delta-u} gives
$\dk=d\epsk/U+O(t^3/U^3)$ and hence
$w_p=1/2-\dk$ and $w_h=1/2+\dk$. Dividing the weighted moments by their
respective weights and expanding to first order in $t/U$ gives
\begin{equation}
\begin{aligned}
E_p
&=\frac{U/4+\epsk/2}{1/2-\dk}
=\frac U2+(1+d)\epsk+O(t^2/U),\\
E_h
&=\frac{-U/4+\epsk/2}{1/2+\dk}
=-\frac U2+(1+d)\epsk+O(t^2/U).
\end{aligned}
\label{eq:s-normalized-moments-u}
\end{equation}
The atomic contributions $\pm U/2$ therefore cancel in
$(E_p+E_h)/2$, while the same residue imbalance shifts both normalized
sector centers by $U\dk=d\epsk$. This yields
\begin{equation}
\centermom=(1/2+2\Cnn)\epsk+O(t^2/U).
\label{eq:s-center-u}
\end{equation}

\paragraph{From Eq.~\ref{eq:s-center-u} to the GZ.}
The GZ contains both the average moment center and the
weight-imbalance term. Substituting Eqs.~\ref{eq:s-weight-u} and
\ref{eq:s-center-u} into Eq.~\ref{eq:s-common-zero} shows their
cancellation explicitly:
\begin{equation}
\boxed{
\begin{aligned}
\wzero^{(U)}(\mathbf{k})
&=\left[(1/2+2\Cnn)+(2\Cnn-1/2)\right]\epsk+O(t^2/U)\\
&=4\Cnn\epsk+O(t^2/U).
\end{aligned}}
\label{eq:s-u-result}
\end{equation}
The spin-independent $+\epsk/2$ contribution to the moment center is
cancelled by the $-\epsk/2$ contribution from $\dk\sep$. In contrast, the
two spin-correlation terms, each $2\Cnn\epsk$, add. The factor near 1.4 is
therefore not a renormalized hopping chosen from the GZ curve; it is fixed
by two complementary first-moment contributions governed by the same bond
spin correlation.

At the resolution used for the Green functions, the periodic spin background has energy $E_0\simeq-11.229$ for 16 sites, giving
\begin{equation}
\Cnn\simeq-0.351,\qquad s_{\mathrm{NN}}=4\Cnn\simeq-1.404.
\label{eq:s-cnn-resolution}
\end{equation}
This value comes from a single periodic $4\times4$ Heisenberg exact
diagonalization, not from averaging over the continuous mixed-momentum
sectors.  At order $t^2/U$, every such sector produces the same exchange
$J_{ij}=4|t_{ij}|^2/U$: the sector-dependent hopping phases cancel between
the forward and reverse virtual hops.  An average would therefore return
the identical correlator in the $U\to\infty$ limit.
Table~\ref{tab:u-decomposition} verifies the approach of the directly extracted root to this first-moment result.

\begin{table}[H]
\centering
\caption{Mott-side numerical decomposition. Signs are NN coefficients multiplying $\epsk$.}
\label{tab:u-decomposition}
\begin{tabular}{cccc}
\toprule
$U/t$ & direct root & first-moment GZ & $s_{\mathrm{NN}}=4\Cnn$ \\
\midrule
8  & $-1.331$ & $-1.435$ & $-1.404$ \\
12 & $-1.303$ & $-1.415$ & $-1.404$ \\
20 & $-1.363$ & $-1.406$ & $-1.404$ \\
40 & $-1.393$ & $-1.404$ & $-1.404$ \\
\bottomrule
\end{tabular}
\end{table}

For the thermodynamic square-lattice spin-$1/2$ Heisenberg antiferromagnet, the high-precision ground-state energy is $e_0\simeq-0.669$ for $J=1$ \cite{sandvik2026highprecision}. Because there are two nearest-neighbor bonds per site,
\begin{equation}
\Cnn^{\infty}=e_0/2\simeq-0.335,\qquad
s_{\mathrm{NN}}^{\infty}=4\Cnn^{\infty}=2e_0\simeq-1.339.
\label{eq:s-cnn-infty}
\end{equation}
This thermodynamic spin-background value is distinct from the finite
momentum-mixing-resolution correlator that determines the calculated
asymptote.  We expect the latter to approach the thermodynamic value as the
momentum-mixing resolution is increased.

\subsection{Extension to arbitrary hopping classes}
The preceding calculation does not rely on the bond being nearest neighbor.
Resolve the kinetic energy into symmetry-equivalent hopping classes $b$,
\begin{equation}
 H_t=\sum_b H_{t_b},\qquad
 \varepsilon_b(\mathbf k)=-t_b\sum_{\boldsymbol\delta\in b}
 e^{i\mathbf k\cdot\boldsymbol\delta},\qquad
 C_b=\langle\mathbf S_i\cdot\mathbf S_{i+\boldsymbol\delta}\rangle_{\boldsymbol\delta\in b}.
\label{eq:s-shell-definitions}
\end{equation}
Applying the Schrieffer--Wolff and Hellmann--Feynman steps above to one bond
in class $b$ gives
\begin{equation}
 g_b\equiv\langle c^\dagger_{i\sigma}
 c_{i+\boldsymbol\delta,\sigma}\rangle
 =\frac{2t_b}{U}\left(\frac14-C_b\right)
 +O(t_{\max}^2/U^2).
\label{eq:s-shell-density-matrix}
\end{equation}
The coordination of class $b$ enters through the displacement sum in
$\varepsilon_b$; it is not an additional prefactor. Fourier transformation
of Eq.~\ref{eq:s-shell-density-matrix} therefore gives the particle--hole
weight imbalance
\begin{equation}
 \dk=n_{\mathbf k\sigma}-\frac12
 =\frac1U\sum_b\left(2C_b-\frac12\right)
 \varepsilon_b(\mathbf k)+O(t_{\max}^2/U^2).
\label{eq:s-shell-weight-imbalance}
\end{equation}

To obtain the moment center without assuming that particle--hole symmetry
divides the dispersive first moment equally, write
\begin{equation}
 M_p=\frac U4+m_p,\qquad M_h=-\frac U4+m_h,
 \qquad m_p+m_h=\sum_b\varepsilon_b(\mathbf k),
\label{eq:s-shell-weighted-moments}
\end{equation}
where a momentum-independent midgap shift has been removed. Using
$w_p=1/2-\dk$ and $w_h=1/2+\dk$, expansion of the normalized moments gives
\begin{align}
 E_p&=\frac{U/4+m_p}{1/2-\dk}
     =\frac U2+2m_p+U\dk+O(t_{\max}^2/U),\nonumber\\
 E_h&=\frac{-U/4+m_h}{1/2+\dk}
     =-\frac U2+2m_h+U\dk+O(t_{\max}^2/U).
\label{eq:s-shell-normalized-moments}
\end{align}
Consequently,
\begin{equation}
 \centermom=\frac{E_p+E_h}{2}
 =\sum_b\varepsilon_b(\mathbf k)+U\dk+O(t_{\max}^2/U),
 \qquad \sep=U+O(t_{\max}).
\label{eq:s-shell-center-separation}
\end{equation}
Substitution of Eqs.~\ref{eq:s-shell-weight-imbalance} and
\ref{eq:s-shell-center-separation} into the shared GZ identity,
Eq.~\ref{eq:s-common-zero}, exposes the cancellation for each hopping class:
\begin{align}
 \wzero(\mathbf k)
 &=\sum_b\varepsilon_b(\mathbf k)+2U\dk
   +O(t_{\max}^2/U)\nonumber\\
 &=\sum_b\left[1+2\left(2C_b-\frac12\right)\right]
   \varepsilon_b(\mathbf k)+O(t_{\max}^2/U)\nonumber\\
 &=4\sum_b C_b\varepsilon_b(\mathbf k)+O(t_{\max}^2/U).
\label{eq:s-hopping-channel-result}
\end{align}
Thus, at leading order,
\begin{equation}
 \boxed{\wzero(\mathbf k)=4\sum_b C_b\varepsilon_b(\mathbf k)}.
\label{eq:s-hopping-channel-boxed}
\end{equation}
The two terms proportional to $2C_b$ add, while the spin-independent
$+\varepsilon_b/2$ and $-\varepsilon_b/2$ pieces cancel. This is the origin
of the common factor $4$. It is unrelated to coordination: a third-neighbor
hopping class contributes $4C_{\mathrm{3N}}\varepsilon_{\mathrm{3N}}$, even when
its displacement set contains a different number of bonds.

\section{CDW-side derivation and hopping-channel charge rule}
For $U=0$ and $V\gg t$, the half-filled atomic background alternates doubly occupied and empty sites. Move one electron from a doubly occupied source to a neighboring empty target. In the intermediate configuration, interactions with the three other occupied neighbors of the target cost $3(2V)$, and the source--target bond contributes $V$. The virtual hopping denominator is therefore
\begin{equation}
\Delta_{\mathrm{hop}}=3(2V)+V=7V.
\label{eq:s-seven-v}
\end{equation}
To make explicit why no additional many-body correlator multiplies this
denominator, choose one of the two atomic checkerboard product states
$|C\rangle$ and orient a nearest-neighbor bond from a doubly occupied site
$i\in A$ to an empty site $j\in B$. For a fixed spin, define the normalized
one-hop defect $|b_{ij,\sigma}\rangle=c^\dagger_{j\sigma}c_{i\sigma}|C\rangle$.
To apply first-order perturbation theory explicitly, split
$H=H_0+H_t$, where $H_0$ contains the atomic density interaction and
$H_t=-t\sum_{\langle lm\rangle\tau}
(c^\dagger_{l\tau}c_{m\tau}+\mathrm{H.c.})$.  The defect is an eigenstate of
$H_0$ with $E_b-E_C=7V$.  The term
$-t c^\dagger_{j\sigma}c_{i\sigma}$ in $H_t$ gives
$\langle b_{ij,\sigma}|H_t|C\rangle=-t$, so its Rayleigh--Schr\"odinger
coefficient is
\[
 \frac{\langle b_{ij,\sigma}|H_t|C\rangle}{E_C-E_b}
 =\frac{-t}{-7V}=\frac{t}{7V}.
\]
The distinct one-hop Fock states are orthogonal.  Moreover, $H_t$ has no
matrix element between the two atomic checkerboard states at first order,
so choosing either state is sufficient for this local calculation.  The
standard first-order wave function and its bond coherence are therefore
\begin{align}
\langle b_{ij,\sigma}|b_{ij,\sigma}\rangle
 &=\langle C|n_{i\sigma}(1-n_{j\sigma})|C\rangle=1,
 \nonumber\\
|\Psi_C\rangle
 &=|C\rangle+
 \sum_{b}\frac{\langle b|H_t|C\rangle}{E_C-E_b}|b\rangle
 +O(t^2/V^2)\nonumber\\
 &=|C\rangle+\frac{t}{7V}
 \sum_{\substack{\langle i\in A,j\in B\rangle\\ \sigma}}
 |b_{ij,\sigma}\rangle+O(t^2/V^2),
 \nonumber\\
\langle\Psi_C|c^\dagger_{i\sigma}c_{j\sigma}|\Psi_C\rangle
 &=\frac{t}{7V}\left[
 \langle C|c^\dagger_{i\sigma}c_{j\sigma}|b_{ij,\sigma}\rangle
 +\langle b_{ij,\sigma}|c^\dagger_{i\sigma}c_{j\sigma}|C\rangle
 \right]+O(t^3/V^3)\nonumber\\
 &=\frac{t}{7V}
 \langle C|c^\dagger_{i\sigma}c_{j\sigma}
 c^\dagger_{j\sigma}c_{i\sigma}|C\rangle+O(t^3/V^3)
 \nonumber\\
 &=\frac{t}{7V}
 \langle C|n_{i\sigma}(1-n_{j\sigma})|C\rangle+O(t^3/V^3)
 =\frac{t}{7V}+O(t^3/V^3).
\label{eq:s-density-matrix-v}
\end{align}
The second cross term in the square brackets vanishes because
$c_{j\sigma}|C\rangle=0$, whereas the first reverses precisely the hop that
created $|b_{ij,\sigma}\rangle$.  The norm of $|\Psi_C\rangle$ differs from
unity only at $O(t^2/V^2)$, so normalization changes this $O(t/V)$
coherence first at $O(t^3/V^3)$.
The factor multiplying $t/(7V)$ is therefore exactly one: in the atomic
checkerboard state $n_{i\sigma}=1$ and $n_{j\sigma}=0$ for either spin.
Both an empty site and an on-site doublon carry zero spin, so there is no
degenerate bond-spin background whose overlap must be averaged. This is the
essential difference from the Mott state, where two singly occupied sites
have a fluctuating singlet--triplet composition and the allowed virtual-hop
weight is the singlet projector, producing
$1/4-\langle\mathbf S_i\!\cdot\!\mathbf S_j\rangle$. Here the doublon--empty
bond fixes the spin matrix element kinematically; the two spin channels give
the same unit factor (their spin-summed coherence is $2t/(7V)$). Spin
correlations induced in the virtual defect have weight $O(t^2/V^2)$ and do
not modify the leading $O(t/V)$ coherence. Moreover, the bipartite gauge
transformation that reverses the sign of $t$ also reverses this bond
coherence, so only odd powers occur and the next correction is
$O(t^3/V^3)$. The translated checkerboard product state gives the same local
matrix element, as does the translation-symmetric combination of the two used
in the calculation.

The momentum dependence follows directly by Fourier transforming this bond
coherence.  The on-site contribution remains
$\langle n_{i\sigma}\rangle=1/2$, while the four directed NN displacements
give
\begin{equation}
\begin{aligned}
n_{\mathbf k\sigma}
&=\frac12+2g_{\mathrm{NN}}(\cos k_x+\cos k_y)+O(t^3/V^3)\\
&=\frac12-\frac{1}{7V}\epsk+O(t^3/V^3),
\end{aligned}
\label{eq:s-nk-v}
\end{equation}
where $g_{\mathrm{NN}}=t/(7V)+O(t^3/V^3)$ from
Eq.~\ref{eq:s-density-matrix-v} and
$\epsk=-2t(\cos k_x+\cos k_y)$.  The zeroth-moment identity therefore gives
\begin{equation}
\dk=n_{\mathbf{k}\sigma}-\frac12
=-\frac{1}{7V}\epsk+O(t^3/V^3).
\label{eq:s-delta-v}
\end{equation}

It remains to identify which term of Eq.~\ref{eq:s-common-zero} survives at
$O(t)$.  At particle--hole symmetry, $\mu=4V$, the atomic addition and removal
centers are $+4V$ and $-4V$, so their separation retains the atomic value
$\sep=8V+O(t)$.  Particle--hole symmetry by itself maps addition at
$\mathbf k$ to removal at $\mathbf Q-\mathbf k$ and therefore does not fix the
common center at the same momentum.  We instead use the exact extended-Hubbard
first-moment identity derived in Sec.~\ref{sec:s-finite-uv-moments}.  At $U=0$
it reads $\mathcal M_1=M_p+M_h=\epsk-V\mathcal F_{\mathbf k}$, while
$\mathcal M_1=\centermom-\dk\sep$ algebraically.  To the order retained here,
$\mathcal F_{\mathbf k}=\dk=-\epsk/(7V)$.  Hence
\begin{equation}
\begin{aligned}
 \centermom
 &=\mathcal M_1+\dk\sep
 =\epsk-V\dk+8V\dk+O(t^2/V)\\
 &=\epsk+7V\dk+O(t^2/V)=O(t^2/V),
 \qquad \sep=8V+O(t).
\end{aligned}
\label{eq:s-separation-v}
\end{equation}
The two $O(t)$ terms cancel exactly in the complete-sector first moment.  Thus
the leading GZ comes entirely from $\dk\sep$; unlike the Mott case, there is no
second $O(t)$ contribution from $\centermom$.  This statement is specific to
NN hopping at $t'=0$: a same-sublattice NNN hop can propagate the pre-existing
charge defect and therefore enters $\centermom$ at first order, as derived
below.
Substitution into Eq.~\ref{eq:s-common-zero} gives
\begin{equation}
\boxed{\wzero^{(V)}(\mathbf{k})=-\frac{1}{7V}\epsk(8V)+O(t^2/V)
=-\frac87\epsk+O(t^2/V).}
\label{eq:s-v-result}
\end{equation}
The calculated $V$-axis sequence converges monotonically to this parameter-free value.

The local argument generalizes to a bipartite lattice of coordination $z$. The virtual defect costs $(2z-1)V$, whereas the addition--removal separation is $2zV$, yielding
\begin{equation}
\boxed{\wzero^{(V)}(\mathbf{k})
=-\frac{2z}{2z-1}\epsk+O(t^2/V).}
\label{eq:s-v-general-z}
\end{equation}
Unlike the Mott-side coefficient, the leading CDW-side number depends only on local coordination and does not require a nontrivial thermodynamic correlation estimate.

\subsection{Charge rule for separate hopping channels and numerical check}
\label{sec:s-charge-bond-rule}
For nonzero next-nearest-neighbor hopping, write the bare dispersion as
\begin{equation}
 \varepsilon_{\mathrm{NN}}(\mathbf k)=-2t(\cos k_x+\cos k_y),\qquad
 \varepsilon_{\mathrm{NNN}}(\mathbf k)=-4\tp\cos k_x\cos k_y.
 \label{eq:s-charge-components}
\end{equation}
The two hopping contributions act differently in the atomic checkerboard background. The NN operator connects opposite sublattices and creates the $7V$ defect counted in Eq.~\ref{eq:s-seven-v}; combining its weight imbalance with the $8V$ addition--removal separation gives the coefficient $-8/7$.

A NNN hop instead connects sites on the same checkerboard sublattice. Let $|\mathbf R,+\rangle$ denote a particle-addition defect on the empty sublattice and $|\mathbf R,-\rangle$ a particle-removal defect on the occupied sublattice, and let $P_\pm$ project onto the corresponding one-defect manifolds. To first order in $\tp$, projection of the hopping operator gives
\begin{equation}
\begin{aligned}
 P_+H_{\tp}P_+
 &=\sum_{\mathbf k}\varepsilon_{\mathrm{NNN}}(\mathbf k)
 |\mathbf k,+\rangle\langle\mathbf k,+|,\\
 -P_-H_{\tp}P_-
 &=\sum_{\mathbf k}\varepsilon_{\mathrm{NNN}}(\mathbf k)
 |\mathbf k,-\rangle\langle\mathbf k,-|.
\end{aligned}
\label{eq:s-projected-nnn-defect}
\end{equation}
The minus sign in the second line is the signed particle-removal convention $E_h<0$. Thus particle addition and removal acquire the same $\varepsilon_{\mathrm{NNN}}$ contribution to their signed moment centers. The NNN term propagates a defect already present in the Lehmann sector, so it carries neither the $7V$ virtual denominator nor a spin-overlap factor. It enters the common moment center, while the leading NN weight imbalance is unchanged. Consequently,
\begin{equation}
 \boxed{\wzero^{(V)}(\mathbf k)=-\frac87\varepsilon_{\mathrm{NN}}(\mathbf k)
 +\varepsilon_{\mathrm{NNN}}(\mathbf k)+O(t_{\max}^2/V).}
 \label{eq:s-hopping-channel-charge}
\end{equation}
Frequencies are measured from the chemical potential. At nonzero $\tp$, where particle--hole symmetry no longer fixes the midpoint, a momentum-independent midgap shift is removed before the two hopping-channel coefficients are extracted; any remaining offset belongs to the indicated subleading correction.

Table~\ref{tab:s-hopping-channel-tests} compares the coefficients obtained from the 120 unique path momenta with the independent strong-coupling predictions. The large residual of a single scale multiplying the complete bare $t$--$\tp$ band is visible directly in the curves; resolving the NN and NNN contributions removes that shape mismatch.
\begin{table}[H]
\centering
\caption{\textbf{GZ coefficients for separate hopping channels at $\eta=0.05$.} The zero-offset form is $\wzero=s_{\mathrm{NN}}\varepsilon_{\mathrm{NN}}+s_{\mathrm{NNN}}\varepsilon_{\mathrm{NNN}}$. Parenthetical entries are parameter-free strong-coupling predictions. The final column is the residual root RMSE of the two-channel description.}
\label{tab:s-hopping-channel-tests}
\small
\begin{tabular}{ccccc}
\toprule
Mechanism & $\tp/t$ & $s_{\mathrm{NN}}$ & $s_{\mathrm{NNN}}$ & RMSE$/t$ \\
\midrule
Mott & $0$     & $-1.393$ ($-1.404$) & ---                & 0.028 \\
Mott & $-0.25$ & $-1.391$ ($-1.403$) & 0.834 (0.836)      & 0.030 \\
Mott & $-0.4$  & $-1.384$ ($-1.398$) & 0.793 (0.795)      & 0.036 \\
\addlinespace
CDW  & $0$     & $-1.142$ ($-1.143$) & ---                & $3\times10^{-4}$ \\
CDW  & $-0.25$ & $-1.142$ ($-1.143$) & 1.000 (1.000)      & $3\times10^{-4}$ \\
CDW  & $-0.4$  & $-1.142$ ($-1.143$) & 1.000 (1.000)      & $3\times10^{-4}$ \\
\bottomrule
\end{tabular}
\end{table}

Each spectral group contains 121 plotted $\Gamma$--$X$--$M$--$\Gamma$ rows, with the final $\Gamma$ repeating the first; coefficient extraction uses the 120 unique momenta. At each momentum we enumerate all sign changes of $\ReG$ inside the central gap and track the continuous branch with low spectral weight. The six interaction-axis groups in Table~\ref{tab:s-hopping-channel-tests} have a root at every unique momentum. In the finite-$(U,V)$ scans below, unavailable rows are omitted from coefficient extraction and are never interpolated into a fit.

\section{Finite-interaction spectral reorganization and full scans}
\begin{figure}[H]
\centering
\includegraphics[width=0.96\linewidth]{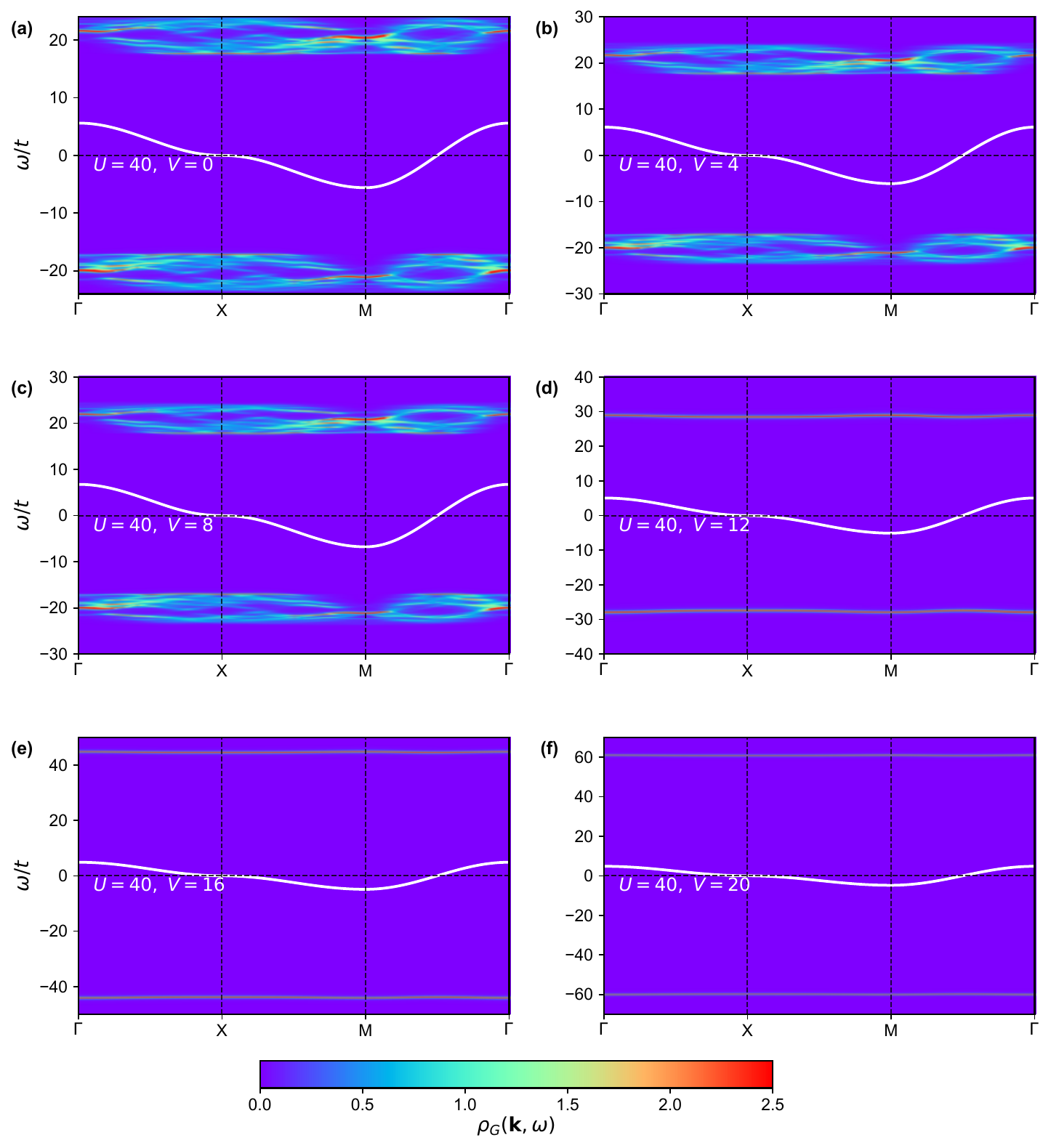}
\caption{\textbf{Finite-interaction reorganization of the GZ dispersion.} Spectral functions at fixed $U=40$ for (a) $V=0$, (b) $V=4$, (c) $V=8$, (d) $V=12$, (e) $V=16$, and (f) $V=20$. White curves are the extracted gap-interior GZ branches. The coefficient magnitude first increases on the Mott side and then drops onto the CDW-side branch. One unavailable momentum row in panels (d)--(f) is interpolated only for image display; all quantitative fits omit unavailable rows.}
\label{fig:s-finite-spectra}
\end{figure}

Figure~\ref{fig:s-finite-spectra} gives the spectral sequence underlying the finite-interaction discussion in the main text. At fixed $U=40$, the denser progression $V/t=0,4,8,12,16,$ and 20 follows the growth of the Mott-side GZ dispersion and its transfer to the CDW-side branch while the central gap remains open. The $V=12$, 16, and 20 source sets each lack one momentum row. That row is interpolated between its neighbors only in the displayed raster and white guide; every quantitative root fit omits unavailable data.

\begin{figure}[H]
\centering
\includegraphics[width=0.98\linewidth]{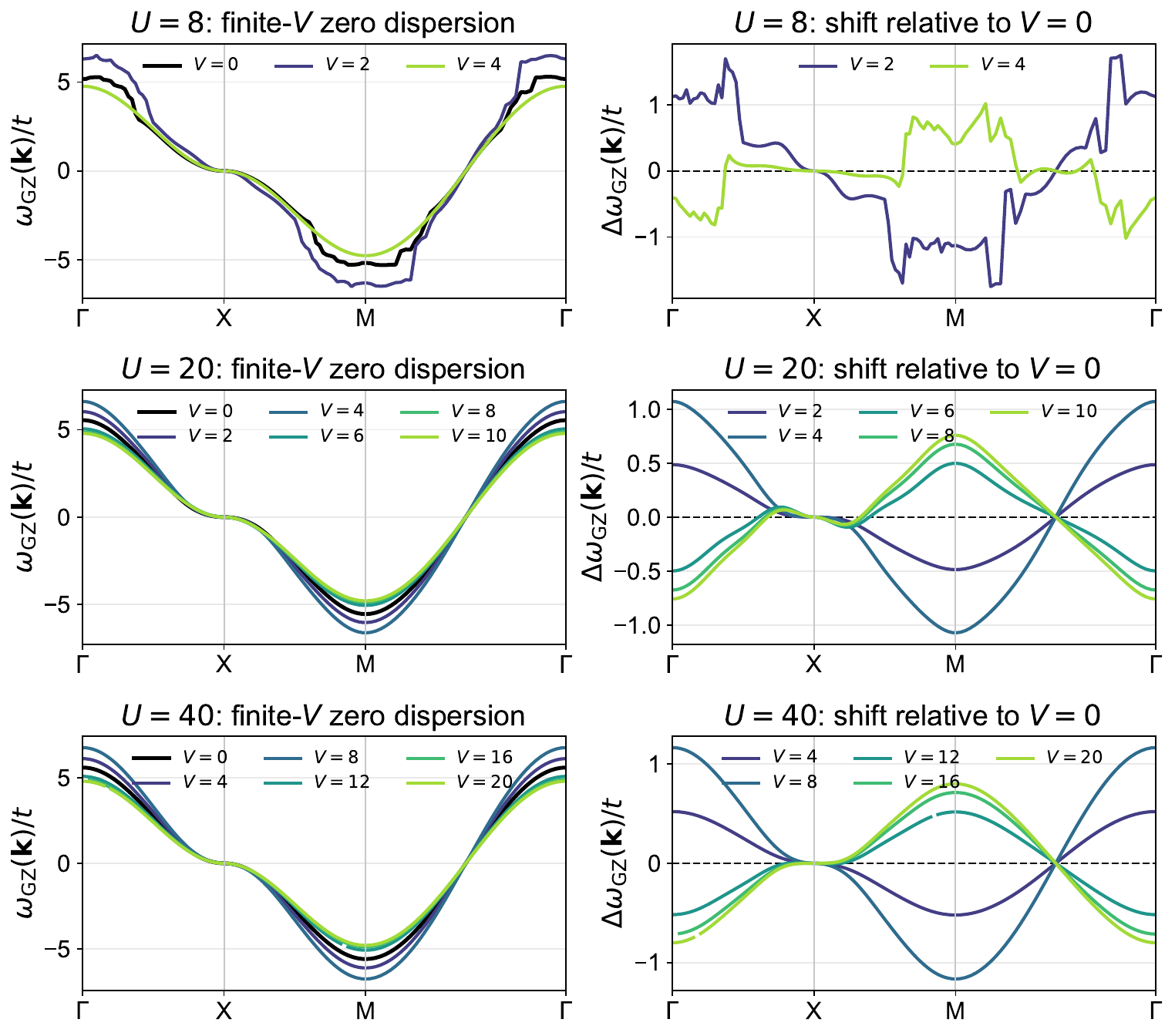}
\caption{\textbf{Extended finite-$V$ scans.} GZ dispersions at fixed $U$ (left) and their changes from $V=0$ (right).}
\label{fig:s-vscan}
\end{figure}

Figure~\ref{fig:s-vscan} displays the broader finite-$V$ sequence behind Fig.~\ref{fig:s-finite-spectra}. Even-integer $V$ values are shown for $U=8$ and 20; for $U=40$, the displayed sequence uses multiples of four. The left column gives $\wzero(\mathbf{k})$ and the right column subtracts the corresponding $V=0$ curve. The change is not a uniform shift: the dispersion grows on the Mott side and changes branch on the CDW side.

\section{Finite-$(U,V)$ strong-coupling laws and numerical validation}
\label{sec:s-finite-uv-laws}

\subsection{Finite-$(U,V)$ Mott branch}
\label{sec:s-finite-uv-mott}
In the atomic Mott background every site is singly occupied. A NN hop creates
a doublon--holon pair. The on-site energy increases by $U$, while the
interaction energy of the hopped bond decreases from $V$ to zero; changes on
the other six bonds cancel pairwise. The neutral virtual-defect energy cost is
therefore
\begin{equation}
 \Delta E_{\mathrm M}=U-V .
 \label{eq:s-mott-denominator-uv}
\end{equation}
The corresponding leading spin Hamiltonian is
\begin{equation}
 H_{\rm spin}
 =\frac{4t^2}{U-V}\sum_{\langle ij\rangle}
 \left(\mathbf S_i\cdot\mathbf S_j-\frac14\right)
 +O\!\left(t^3/(\Delta E_{\mathrm M})^2\right).
 \label{eq:s-spin-hamiltonian-uv}
\end{equation}
Hellmann--Feynman differentiation with respect to one bond hopping gives
\begin{equation}
 g_{\rm M}\equiv
 \langle c^\dagger_{i\sigma}c_{j\sigma}\rangle
 =\frac{2t}{U-V}\left(\frac14-\Cnn\right)
 +O\!\left(t^2/(\Delta E_{\mathrm M})^2\right).
 \label{eq:s-mott-coherence-uv}
\end{equation}
Writing $d=2\Cnn-1/2$ and Fourier transforming yields
\begin{equation}
 \mathcal F_{\mathbf k}=\dk
 =\frac{d}{U-V}\epsk+O\!\left(t^2/(\Delta E_{\mathrm M})^2\right).
 \label{eq:s-mott-delta-uv}
\end{equation}

Although the virtual hopping cost is $\Delta E_{\mathrm M}=U-V$, the atomic addition and
removal poles remain at $+U/2$ and $-U/2$. The $4V$ change in the local bond
energy is cancelled by the $4V$ part of $\mu=U/2+4V$, so
\begin{equation}
 \Delta_{\rm M}^{(0)}=U .
 \label{eq:s-mott-separation-uv}
\end{equation}
Using Eqs.~\ref{eq:s-uv-m1-exact},
\ref{eq:s-mott-delta-uv}, and
\ref{eq:s-mott-separation-uv}, the normalized-sector center is
\begin{align}
 \mathcal C_{\mathbf k}^{\rm M}
 &=\mathcal M_1+\dk\Delta_{\rm M}^{(0)}\nonumber\\
 &=\epsk+(U-V)\dk+O(t^2/\Delta E_{\mathrm M})\nonumber\\
 &=\left(\frac12+2\Cnn\right)\epsk+O(t^2/\Delta E_{\mathrm M}).
 \label{eq:s-mott-center-uv}
\end{align}
The GZ is consequently
\begin{align}
 \wzero^{\rm M}(\mathbf k)
 &=\mathcal M_1+2\dk\Delta_{\rm M}^{(0)}+O(t^2/\Delta E_{\mathrm M})\nonumber\\
 &=\left[1+\frac{2U-V}{U-V}
 \left(2\Cnn-\frac12\right)\right]\epsk+O(t^2/\Delta E_{\mathrm M}),
 \label{eq:s-mott-result-uv}
\end{align}
so that, in the form used in Eq.~(8) of the main text,
\begin{equation}
 \boxed{
 s_{\mathrm{NN}}^{\rm M}(U,V)
 =1+\frac{2U-V}{U-V}
 \left(2\Cnn-\frac12\right)
 +O\!\left(\frac{t}{\Delta E_{\mathrm M}}\right).}
 \label{eq:s-mott-scale-uv}
\end{equation}

\subsection{Finite-$(U,V)$ CDW branch and moment-center cancellation}
\label{sec:s-finite-uv-cdw}
In the atomic checkerboard background, a NN hop takes one electron from a
doubly occupied site to an empty site. The initial on-site energy $U$ is
removed. In the intermediate state the target interacts with its three other
doubly occupied neighbors at cost $3(2V)$, and the source--target bond costs
another $V$. The neutral virtual-defect energy cost is therefore
\begin{equation}
 \Delta E_{\mathrm C}=7V-U .
 \label{eq:s-cdw-denominator-uv}
\end{equation}
First-order perturbation theory gives the directed per-spin coherence
\begin{equation}
 g_{\rm C}=\frac{t}{7V-U}+O\!\left(t^2/(\Delta E_{\mathrm C})^2\right),
 \label{eq:s-cdw-coherence-uv}
\end{equation}
and hence
\begin{equation}
 \mathcal F_{\mathbf k}=\dk
 =-\frac{1}{7V-U}\epsk+O\!\left(t^2/(\Delta E_{\mathrm C})^2\right).
 \label{eq:s-cdw-delta-uv}
\end{equation}

At $\mu=U/2+4V$, adding one electron to an empty checkerboard site costs
$8V-\mu=4V-U/2$. Removing one electron from a doublon costs
$\mu-U=4V-U/2$. The signed atomic sector separation is therefore
\begin{equation}
 \Delta_{\rm C}^{(0)}=8V-U .
 \label{eq:s-cdw-separation-uv}
\end{equation}
The exact first-moment sum rule now gives
\begin{equation}
 \mathcal M_1(\mathbf k)
 =\epsk-V\mathcal F_{\mathbf k}
 =\epsk-V\dk+O(t^2/\Delta E_{\mathrm C})
 =\frac{8V-U}{7V-U}\epsk+O(t^2/\Delta E_{\mathrm C}).
 \label{eq:s-cdw-m1-uv}
\end{equation}
It follows without assuming anything about the individual addition and
removal dispersions at the same momentum that
\begin{align}
 \mathcal C_{\mathbf k}^{\rm C}
 &=\mathcal M_1+\dk\Delta_{\rm C}^{(0)}\nonumber\\
 &=\left[\frac{8V-U}{7V-U}-\frac{8V-U}{7V-U}\right]\epsk
 +O(t^2/\Delta E_{\mathrm C})\nonumber\\
 &=O(t^2/\Delta E_{\mathrm C}).
 \label{eq:s-cdw-center-exact-uv}
\end{align}
This is the required $O(t)$ moment-center cancellation for the complete
particle-addition and particle-removal sectors. Finally,
\begin{align}
 \wzero^{\rm C}(\mathbf k)
 &=\mathcal M_1+2\dk\Delta_{\rm C}^{(0)}+O(t^2/\Delta E_{\mathrm C})\nonumber\\
 &=-\frac{8V-U}{7V-U}\epsk+O(t^2/\Delta E_{\mathrm C}),
 \label{eq:s-cdw-result-uv}
\end{align}
and therefore
\begin{equation}
 \boxed{s_{\mathrm{NN}}^{\rm C}(U,V)
 =-\frac{8V-U}{7V-U}
 +O\!\left(\frac{t}{\Delta E_{\mathrm C}}\right).}
 \label{eq:s-cdw-scale-uv}
\end{equation}
For $U=0$, its leading term reduces to $-8/7$. Denoting the leading rational
term by $s_{\mathrm{NN},0}^{\rm C}$ and writing $x=U/(4V)$,
\begin{equation}
 s_{\mathrm{NN},0}^{\rm C}
 =-\frac{8-4x}{7-4x}
 =-\frac87-\frac{4}{49}x-\frac{16}{343}x^2+O(x^3),
 \label{eq:s-cdw-expansion-uv}
\end{equation}
so the first correction to the $V$-only limit is linear, rather than
quadratic, in $U/(4V)$.

\section{All-momentum static observables and Mott--CDW brackets}
For each $U/t=4,8,\ldots,40$ in main-text Fig.~5(b), we define $V_{\rm Mott}$ as the largest sampled interaction on the Mott branch and $V_{\rm CDW}$ as the smallest sampled interaction on the CDW branch. Both endpoints are evaluated on the complete 861-point symmetry-reduced mixed-momentum grid. Branch assignment follows the coincident discontinuities in the ground-state energy slope, double occupancy, checkerboard structure factor, and nearest-neighbor spin correlation. We plot the midpoint
\begin{equation}
 \left(\frac{V}{U}\right)_c=\frac{V_{\rm Mott}+V_{\rm CDW}}{2U}
\end{equation}
with vertical half-width $(V_{\rm CDW}-V_{\rm Mott})/(2U)$. Thus the bar reports the directly sampled interval containing the branch crossing, rather than a statistical uncertainty or a fit interval.

\begin{table}[H]
\centering
\caption{\textbf{Directly sampled Mott--CDW brackets used in main-text Fig.~5(b).} $V_{\rm Mott}$ is the largest sampled Mott-side value and $V_{\rm CDW}$ is the smallest sampled CDW-side value. The plotted point is their midpoint divided by $U$; the two endpoints, rather than a statistical error model, determine its vertical bar.}
\label{tab:transition-brackets-all-u}
\small
\begin{tabular}{cccc}
\toprule
$U/t$ & $V_{\rm Mott}/t$ & $V_{\rm CDW}/t$ & $(V/U)_c$ \\
\midrule
4  & 1.029  & 1.030  & 0.257 \\
8  & 2.038  & 2.040  & 0.255 \\
12 & 3.032  & 3.034  & 0.253 \\
16 & 4.025  & 4.030  & 0.252 \\
20 & 5.020  & 5.025  & 0.251 \\
24 & 6.020  & 6.025  & 0.251 \\
28 & 7.015  & 7.020  & 0.251 \\
32 & 8.015  & 8.020  & 0.251 \\
36 & 9.010  & 9.020  & 0.250 \\
40 & 10.010 & 10.020 & 0.250 \\
\bottomrule
\end{tabular}
\end{table}

The measured static quantities use the following normalizations, with $N_s=16$ sites and $N_b=2N_s$ distinct NN bonds:
\begin{align}
 D_i&\equiv\frac1{N_s}\sum_i\langle n_{i\uparrow}n_{i\downarrow}\rangle,
 &\frac{K}{N_s}&\equiv\frac{\langle H_{\rm kin}\rangle}{N_s},\nonumber\\
 \langle n_i n_j\rangle_{\rm NN}
 &\equiv\frac1{N_b}\sum_{\langle ij\rangle}\langle n_i n_j\rangle,
 &C_{\rm NN}&\equiv\frac1{N_b}\sum_{\langle ij\rangle}
 \langle\mathbf S_i\!\cdot\!\mathbf S_j\rangle,
 \label{eq:s-static-local-definitions}\\
 S_c(\mathbf Q)&\equiv\frac1{N_s}\sum_{ij}e^{i\mathbf Q\cdot(\mathbf r_i-\mathbf r_j)}
 \langle(n_i-1)(n_j-1)\rangle,
 &S^{zz}(\mathbf Q)&\equiv\frac1{N_s}\sum_{ij}e^{i\mathbf Q\cdot(\mathbf r_i-\mathbf r_j)}
 \langle S_i^zS_j^z\rangle,\nonumber\\
 n(\mathbf k)&\equiv\sum_\sigma\langle c^\dagger_{\mathbf k\sigma}c_{\mathbf k\sigma}\rangle.
 &&
 \label{eq:s-static-structure-definitions}
\end{align}
Thus $n(\mathbf k)$ is spin summed and lies between 0 and 2. The charge-density-wave proxy used below is $m_{\rm CDW}^2=S_c(\pi,\pi)/N_s$.

The all-momentum calculation supplies the order diagnostics that were absent from the Green-function path data. For each $V$, the mixed-momentum grid contains 41 points in each direction. We calculate the wedge $0\le k_x\le k_y\le40$, reflect it under $k_x\leftrightarrow k_y$, and use a symmetry weight of two away from the diagonal. Independent half weights at the $0$ and $\pi$ endpoints implement two-dimensional trapezoidal quadrature. A complete set therefore contains 861 solved wedge points and total quadrature weight 1600. All 14 sets at $V=0,4,8,9,10,10.01,10.02,10.1,10.2,10.4,11,12,16,$ and 20 are complete.

\begin{figure}[H]
\centering
\includegraphics[width=0.98\linewidth]{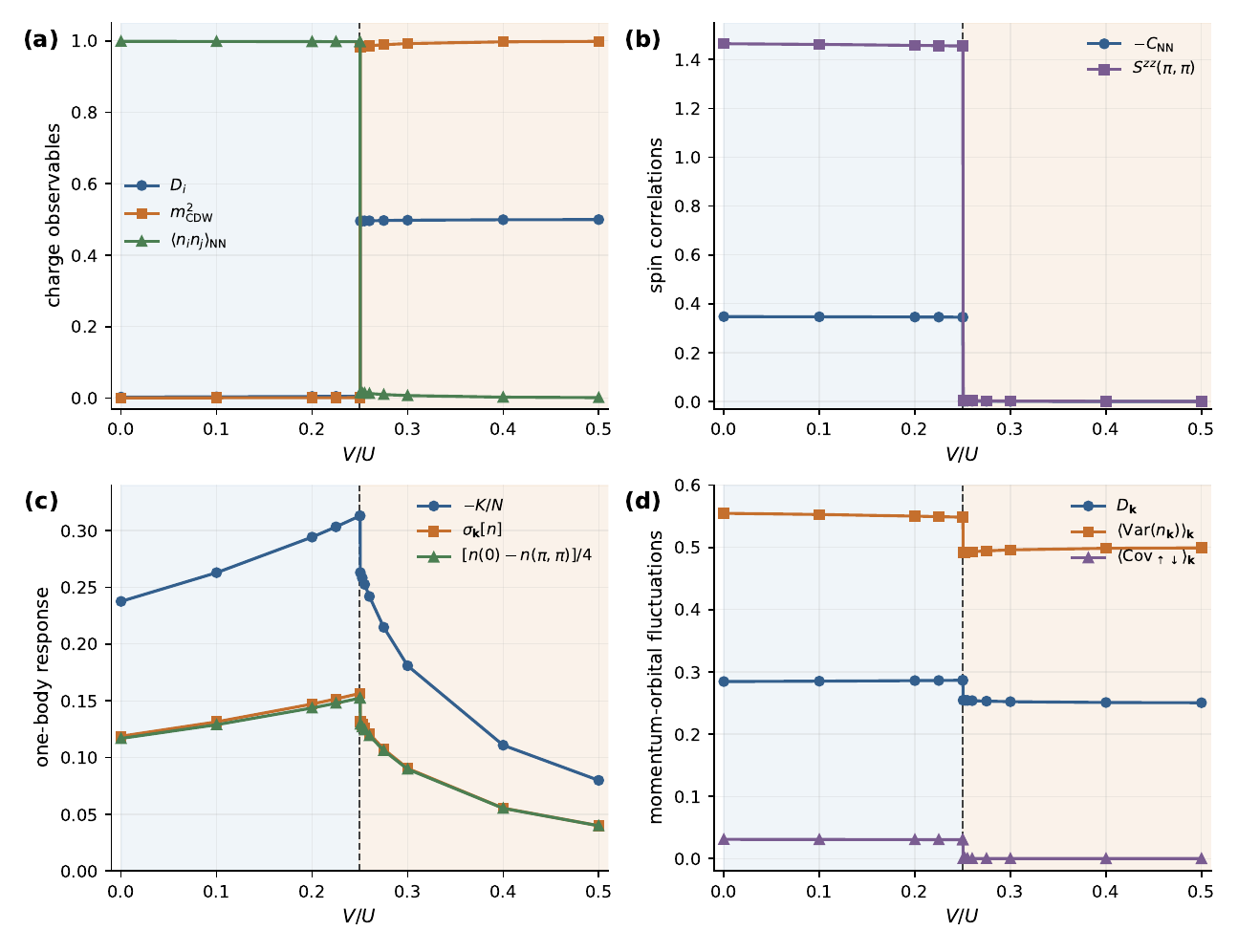}
\caption{\textbf{All-momentum static scan at $U=40$.} The simultaneous changes in charge, spin, kinetic, and momentum observables occur inside the direct bracket $10.01<V_c<10.02$. Momentum-distribution widths measure variation over physical momentum, not statistical uncertainties.}
\label{fig:s-fullobs-static}
\end{figure}

Figure~\ref{fig:s-fullobs-static} shows that charge, spin, kinetic, and momentum observables change branch together. The momentum-to-momentum standard deviations of the local observables are $O(10^{-5})$ near the crossing, so the discontinuity is not produced by a minority of mixed-momentum points. Table~\ref{tab:fullobs-boundary} lists the two directly adjacent points. The exact energy identity
\begin{equation}
 \frac{E_0}{N_s}=\frac{K}{N_s}+UD_i+2V\langle n_i n_j\rangle_{\rm NN}
 \label{eq:s-energy-identity}
\end{equation}
holds over all 12,054 calculated points with maximum residual $2.2\times10^{-10}$.

\begin{table}[H]
\centering
\caption{Directly adjacent points bracketing the branch crossing. Here $m_{\rm CDW}^2=S_c(\pi,\pi)/16$.}
\label{tab:fullobs-boundary}
\begin{tabular}{lcc}
\toprule
Observable & $V=10.01$ & $V=10.02$ \\
\midrule
$D_i$ & 0.005 & 0.496 \\
$\langle n_i n_j\rangle_{\rm NN}$ & 0.997 & 0.015 \\
$m_{\rm CDW}^2$ & 0.001 & 0.984 \\
$C_{\mathrm{NN}}$ & $-0.345$ & $-0.002$ \\
$S^{zz}(\pi,\pi)$ & 1.456 & 0.004 \\
$K/N$ & $-0.313$ & $-0.263$ \\
$n_\Gamma$ & 1.305 & 1.259 \\
$2\langle n_i n_j\rangle_{\rm NN}$ & 1.995 & 0.031 \\
\bottomrule
\end{tabular}
\end{table}

\begin{figure}[H]
\centering
\includegraphics[width=0.99\linewidth]{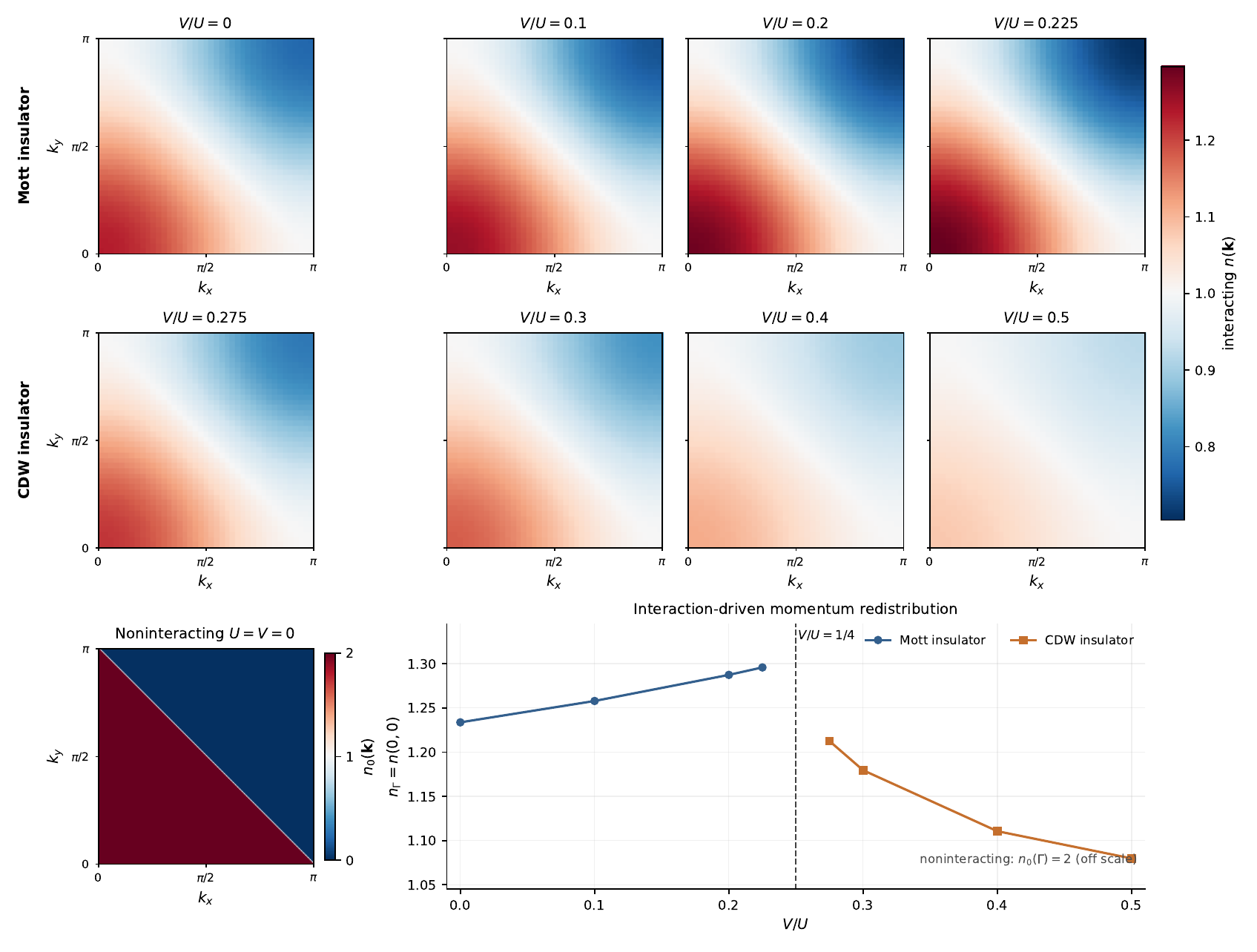}
\caption{\textbf{Two-dimensional momentum occupation across the branch crossing at fixed $U/t=40$.} The first row contains Mott-side points and the second CDW-side points. All interacting panels share one color scale. The third row gives the noninteracting reference, with its separate 0--2 scale, and the measured $n_\Gamma$ evolution. Every displayed interacting map has complete mixed-momentum-grid coverage.}
\label{fig:s-fullobs-nk}
\end{figure}

The full $n(\mathbf{k})$ maps are shown in Fig.~\ref{fig:s-fullobs-nk}. The value $V/U=0.25$ is omitted from the phase-labelled rows because it lies at the atomic crossing; its measured values remain in all quantitative analyses. The noninteracting reference has $n_0(\mathbf{k})=2$ below the Fermi line, zero above it, and one on the line. Interactions redistribute this step into a smooth band-memory profile whose Brillouin-zone average remains one. Particle--hole symmetry is verified numerically through $n(\mathbf{k})+n(\pi-\mathbf{k})=2$ with maximum residual below $7\times10^{-6}$.

\begin{figure}[H]
\centering
\includegraphics[width=0.98\linewidth]{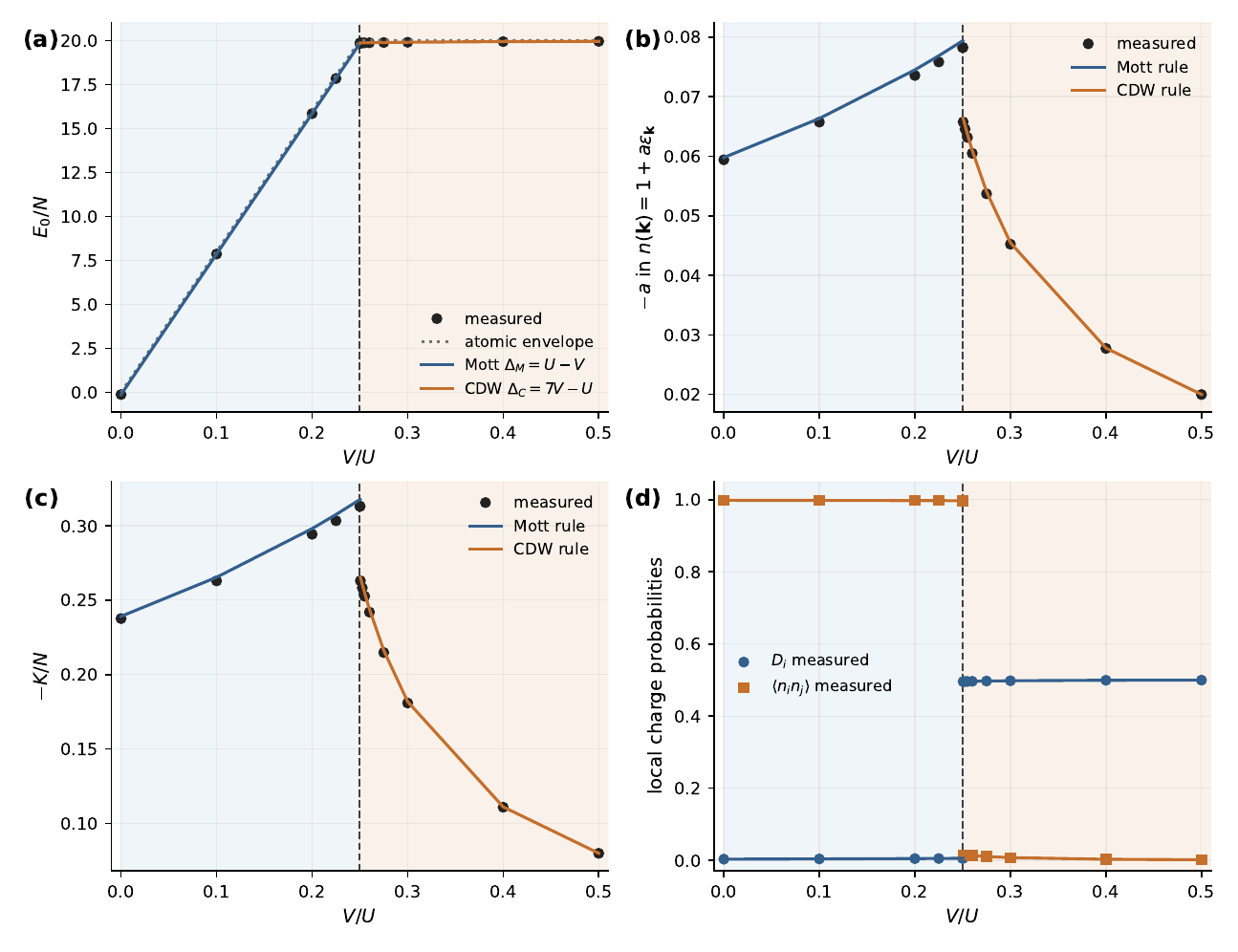}
\caption{\textbf{Strong-coupling validation of the all-momentum observables at fixed $U/t=40$.} Symbols are measured values, and solid curves are the corresponding rows of Eq.~\ref{eq:s-fullobs-strong-rules}: (a) $E_0/N_s$, (b) $-a$ for $n(\mathbf k)=1+a\varepsilon_{\mathbf k}$, (c) $-K/N_s$, and (d) $D_i$ and $\langle n_i n_j\rangle_{\rm NN}$. The dotted curve in (a) is the atomic envelope $V/U=0.25$. Mott- and CDW-side formulas are shown only where their respective atomic backgrounds apply.}
\label{fig:s-fullobs-strong}
\end{figure}

Figure~\ref{fig:s-fullobs-strong} tests the leading strong-coupling description not only for $n(\mathbf{k})$ but also for the energy, kinetic energy, double occupancy, and nearest-neighbor density product. To state explicitly which rule is used in each panel, define
\begin{equation}
 \Delta_{\rm M}=U-V,
 \qquad
 \Delta_{\rm C}=7V-U,
 \qquad
 \mathcal P_{\rm NN}=\frac14-C_{\rm NN},
 \label{eq:s-fullobs-denominators}
\end{equation}
where $\mathcal P_{\rm NN}$ is the nearest-neighbor singlet-projector weight of the Mott background. With $n(\mathbf k)=1+a\varepsilon_{\mathbf k}$, the leading predictions are
\begin{equation}
\begin{array}{c|cc}
 & \text{Mott side} & \text{CDW side} \\
\hline
E_0/N_s
 & 2V-8\mathcal P_{\rm NN}/\Delta_{\rm M}
 & U/2-4/\Delta_{\rm C} \\
a
 & -4\mathcal P_{\rm NN}/\Delta_{\rm M}
 & -2/\Delta_{\rm C} \\
K/N_s
 & -16\mathcal P_{\rm NN}/\Delta_{\rm M}
 & -8/\Delta_{\rm C} \\
D_i
 & 8\mathcal P_{\rm NN}/\Delta_{\rm M}^{2}
 & 1/2-4/\Delta_{\rm C}^{2} \\
\langle n_i n_j\rangle_{\rm NN}
 & 1-4\mathcal P_{\rm NN}/\Delta_{\rm M}^{2}
 & 14/\Delta_{\rm C}^{2}
\end{array}
\label{eq:s-fullobs-strong-rules}
\end{equation}
The five rows of Eq.~\ref{eq:s-fullobs-strong-rules} supply, in order, panel (a), panel (b), panel (c), and the two observables in panel (d); panels (b,c) display $-a$ and $-K/N_s$. Across the applicable points, the fitted $n(\mathbf{k})$ slope differs from the Mott- and CDW-side formulas by at most 1.44\% and 0.68\%, respectively. This agreement identifies virtual hopping as the origin of the momentum differentiation.

\section{Frustrated-hopping robustness}
\begin{figure}[H]
\centering
\includegraphics[width=0.99\linewidth]{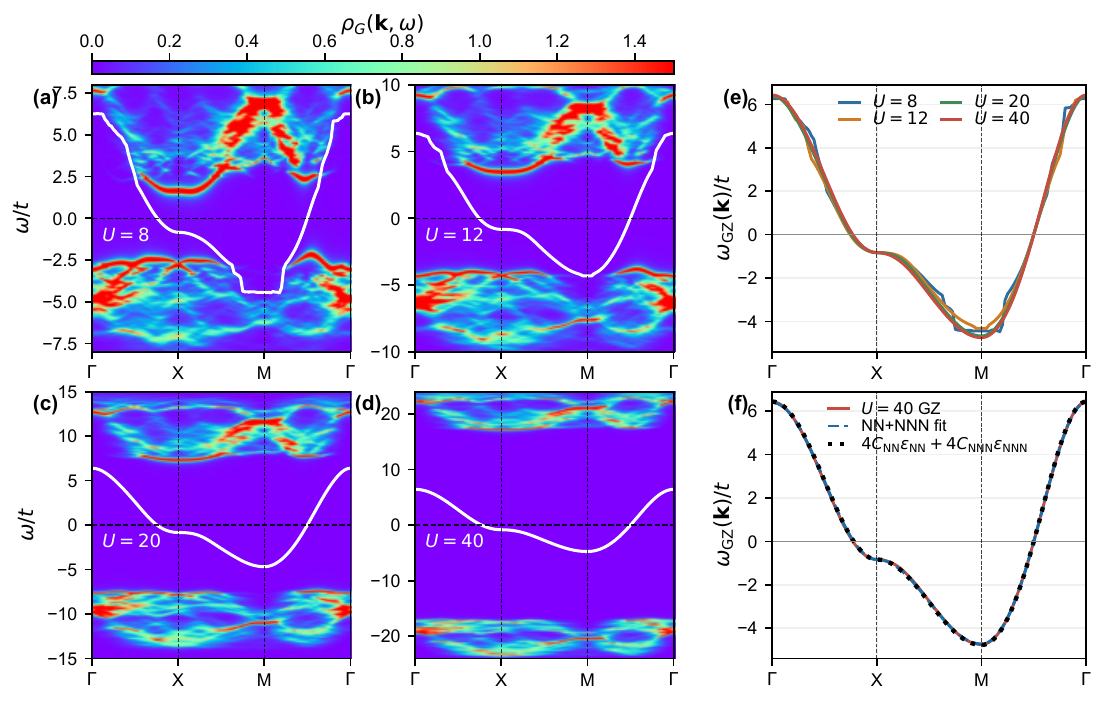}
\caption{\textbf{Hopping-channel GZ dispersion for next-nearest-neighbor hopping.} The layout follows main-text Fig.~3. (a--d) Reproduced spectral maps and GZ dispersions for $\tp=-0.25$, $V=0$, and $U/t=8,12,20,$ and 40. (e) The four extracted GZ dispersions. (f) The $U=40$ curve compared with the independent NN+NNN hopping-channel fit and the parameter-free strong-coupling prediction $s_{\mathrm{NN}}\varepsilon_{\mathrm{NN}}+s_{\mathrm{NNN}}\varepsilon_{\mathrm{NNN}}$, where $s_b=4C_b$.}
\label{fig:s-tprime}
\end{figure}

\begin{figure}[H]
\centering
\includegraphics[width=0.99\linewidth]{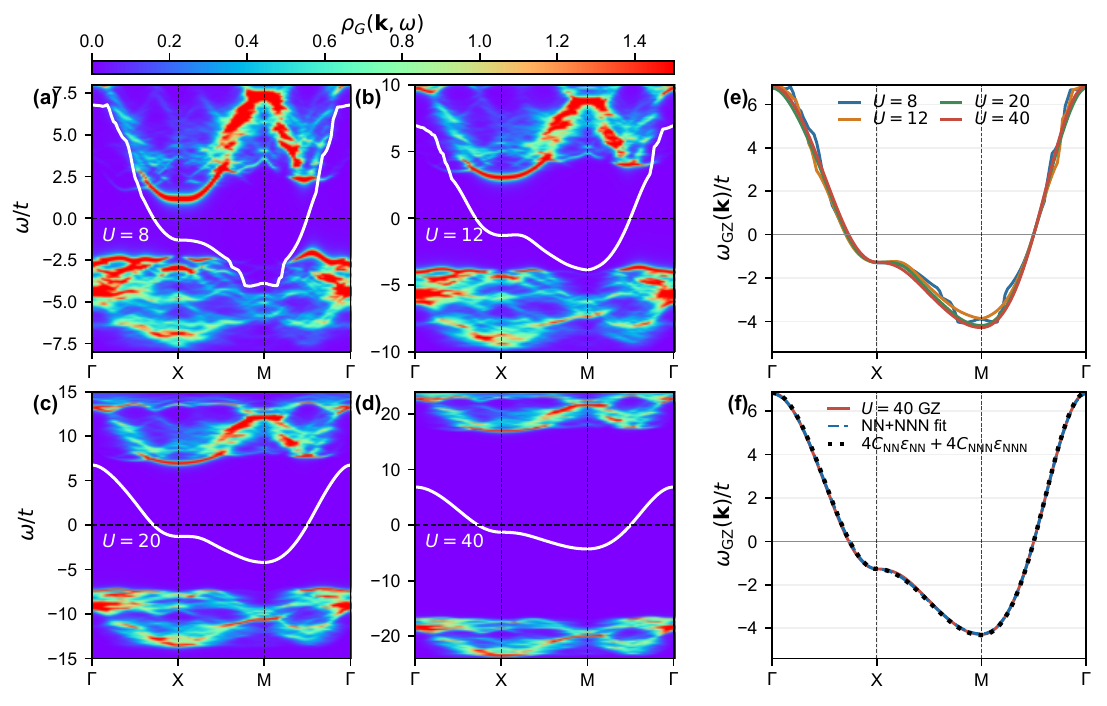}
\caption{\textbf{Hopping-channel GZ dispersion at $\tp=-0.4t$.} The layout follows Fig.~\ref{fig:s-tprime}. (a--d) Reproduced spectral maps and GZ dispersions for $V=0$ and $U/t=8,12,20,$ and 40. (e) The four extracted GZ dispersions. (f) The $U=40$ curve compared with the independent NN+NNN fit and the parameter-free strong-coupling prediction $4C_{\mathrm{NN}}\varepsilon_{\mathrm{NN}}+4C_{\mathrm{NNN}}\varepsilon_{\mathrm{NNN}}$.}
\label{fig:s-tprime-n04}
\end{figure}

\begin{figure}[H]
\centering
\includegraphics[width=0.99\linewidth]{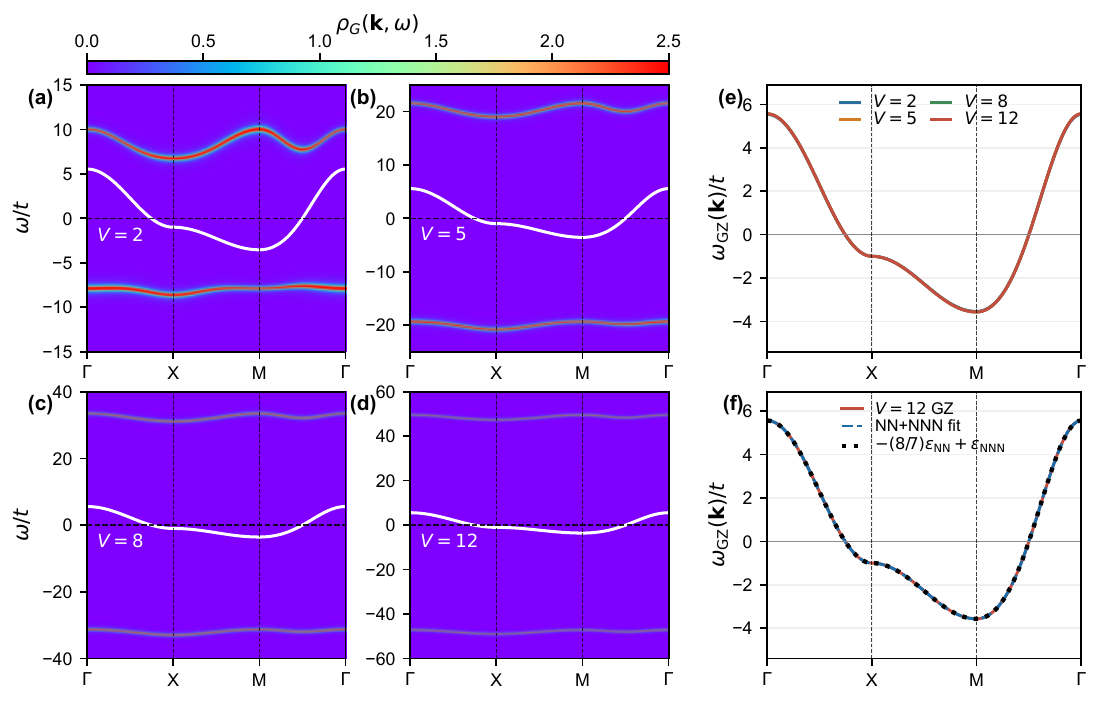}
\caption{\textbf{$V$-only hopping-channel GZ dispersion at $\tp=-0.25t$.} (a--d) Spectral maps and GZ dispersions for $U=0$ and $V/t=2,5,8,$ and 12. (e) The four extracted GZ dispersions. (f) The $V=12$ curve compared with the independent NN+NNN fit and the parameter-free strong-coupling prediction $-(8/7)\varepsilon_{\mathrm{NN}}+\varepsilon_{\mathrm{NNN}}$.}
\label{fig:s-tprime-vonly}
\end{figure}

\begin{figure}[H]
\centering
\includegraphics[width=0.99\linewidth]{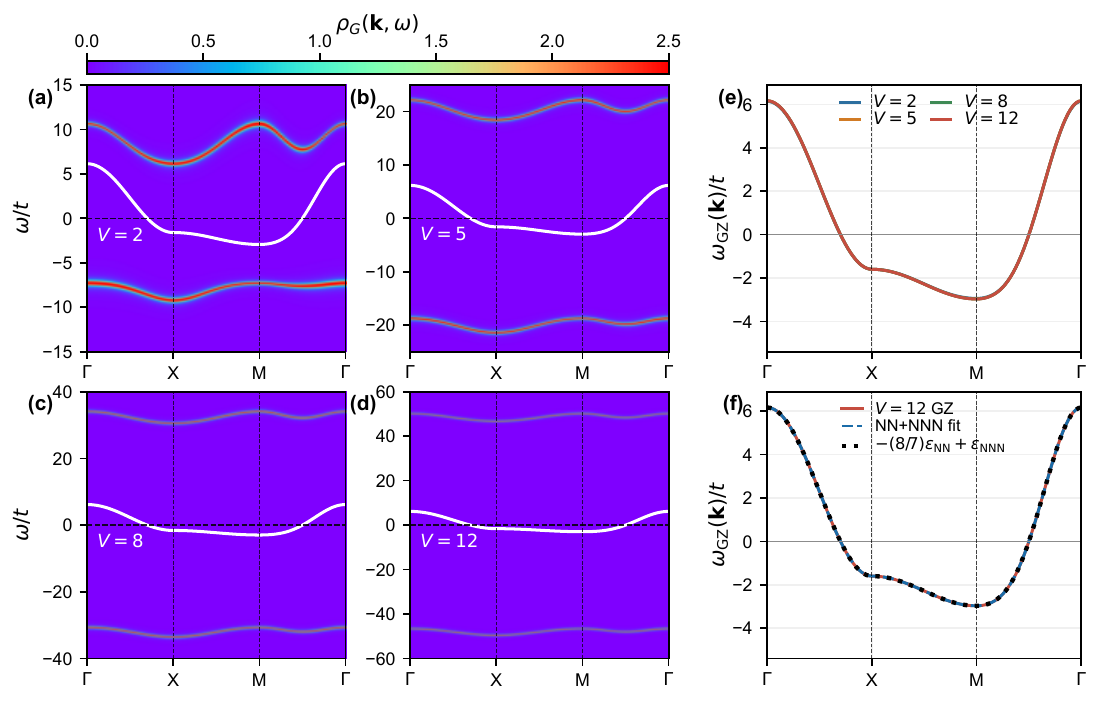}
\caption{\textbf{$V$-only hopping-channel GZ dispersion at $\tp=-0.4t$.} The layout follows Fig.~\ref{fig:s-tprime-vonly}. (a--d) Spectral maps and GZ dispersions for $U=0$ and $V/t=2,5,8,$ and 12. (e) The four extracted GZ dispersions. (f) The $V=12$ curve compared with the independent NN+NNN fit and the parameter-free strong-coupling prediction $-(8/7)\varepsilon_{\mathrm{NN}}+\varepsilon_{\mathrm{NNN}}$.}
\label{fig:s-tprime-vonly-n04}
\end{figure}

The main text presents a compact $\tp=-0.25t$ comparison. Figures~\ref{fig:s-tprime} and \ref{fig:s-tprime-n04} provide the complete $U/t=8,12,20,$ and 40 sequences for $\tp=-0.25t$ and $-0.4t$, respectively, while Figs.~\ref{fig:s-tprime-vonly} and \ref{fig:s-tprime-vonly-n04} give the corresponding $V$-only sequences. A continuous low-weight GZ still traverses the gap, but a single rescaling of the complete bare $t$--$\tp$ band is not the correct strong-coupling description. Separating
\begin{equation}
 \varepsilon_{\mathrm{NN}}(\mathbf k)=-2t(\cos k_x+\cos k_y),\qquad
 \varepsilon_{\mathrm{NNN}}(\mathbf k)=-4\tp\cos k_x\cos k_y,
\label{eq:s-tprime-components}
\end{equation}
the minimal fit is
\begin{equation}
 \wzero(\mathbf k)=s_{\mathrm{NN}}\varepsilon_{\mathrm{NN}}(\mathbf k)
 +s_{\mathrm{NNN}}\varepsilon_{\mathrm{NNN}}(\mathbf k),
 \qquad s_{\mathrm{NN}}<0.
\label{eq:s-tprime-fit}
\end{equation}
The hopping-class moment derivation in Sec.~\ref{sec:s-spin-bond-rule} gives, for the two dispersion components retained here,
\begin{equation}
 \wzero(\mathbf k)=4C_{\mathrm{NN}}\varepsilon_{\mathrm{NN}}(\mathbf k)
 +4C_{\mathrm{NNN}}\varepsilon_{\mathrm{NNN}}(\mathbf k)+O(t_{\max}^2/U),
\label{eq:s-tprime-spin-limit}
\end{equation}
where $C_{\mathrm{NN}}$ and $C_{\mathrm{NNN}}$ are nearest- and next-nearest-neighbor spin correlations. The leading spin Hamiltonian has $J_2/J_1=(\tp/t)^2$. The extracted coefficients and the independently calculated $4C_{\mathrm{NN}}$ and $4C_{\mathrm{NNN}}$ predictions are collected in Table~\ref{tab:s-hopping-channel-tests}. The negative NN coefficient records the inversion caused by $C_{\mathrm{NN}}<0$, whereas the diagonal coefficient remains positive because $C_{\mathrm{NNN}}>0$; the two spin correlations, rather than one global scale, determine the frustrated GZ dispersion.

Figure~\ref{fig:s-tprime-n04} shows that the same separation into NN and NNN contributions remains accurate for the stronger frustration $\tp=-0.4t$ and compares the full GZ dispersion with the corresponding $J_1$--$J_2$ prediction.

The $V$-only sequences in Figs.~\ref{fig:s-tprime-vonly} and \ref{fig:s-tprime-vonly-n04} converge to the distinct CDW-side rule derived in Sec.~\ref{sec:s-charge-bond-rule}: NN hopping carries the local charge-defect coefficient $-8/7$, whereas NNN hopping propagates the defect within one checkerboard sublattice and therefore carries unit coefficient. Figures~\ref{fig:s-tprime-vonly}(a--d) and \ref{fig:s-tprime-vonly-n04}(a--d) make the separation of scales explicit: the particle-addition and particle-removal bands move apart with $V$, whereas the white gap-interior GZ retains an $O(t)$ dispersion. The curves in panels (e) collapse rapidly with increasing $V$, and panels (f) show that at $V/t=12$ the independently fitted NN+NNN form is indistinguishable from $-(8/7)\varepsilon_{\mathrm{NN}}+\varepsilon_{\mathrm{NNN}}$. Increasing $|\tp|$ from $0.25t$ to $0.4t$ changes the GZ shape through the NNN contribution but leaves the NN and NNN coefficients unchanged, confirming that the CDW-side result is controlled by local defect motion rather than by a new global rescaling.

\bibliographystyle{unsrtnat}
\bibliography{references}